\documentclass{article}

\usepackage{cite}
\usepackage{graphicx}
\usepackage{caption}
\usepackage{amssymb}
\usepackage{array}
\usepackage{dcolumn}
\usepackage{bm}
\usepackage{textgreek}
\usepackage{amsmath}
\usepackage[letterpaper, total={7.5in, 10in}]{geometry}
\usepackage{authblk}

\title{Serpentine optical phased arrays for scalable integrated photonic LIDAR beam steering}

\author[1,$\dagger$,*]{Nathan Dostart}
\author[2,$\dagger$]{Bohan Zhang}
\author[2,3]{Anatol Khilo}
\author[1]{Michael Brand}
\author[2]{Kenaish Al Qubaisi}
\author[2]{Deniz Onural}
\author[1]{Daniel Feldkhun}
\author[1]{Kelvin H. Wagner}
\author[2]{Milo\v s A. Popovi\'c}

\affil[1]{Department of Electrical, Computer, and Energy Engineering, University of Colorado, Boulder, CO, 80309, USA}
\affil[2]{Department of Electrical and Computer Engineering, Boston University, Boston, MA, 02215, USA}
\affil[3]{Currently with Ayar Labs, 6460 Hollis St, Ste A, Emeryville, CA, 94608, USA}

\affil[*]{Corresponding author: nathan.dostart@colorado.edu}

\affil[$\dagger$]{These authors contributed equally to this work.}

\date{\today}

\begin{document}
\maketitle

\begin{abstract} 
Optical phased arrays (OPAs) implemented in integrated photonic circuits could enable a variety of 3D sensing, imaging, illumination, and ranging applications, and their convergence in new LIDAR technology.  However, current integrated OPA approaches do not scale -- in control complexity, power consumption, and optical efficiency -- to the large aperture sizes needed to support medium to long range LIDAR.  We present the serpentine optical phased array (SOPA), a new OPA concept that addresses these fundamental challenges and enables architectures that scale up to large apertures.  The SOPA is based on a serially interconnected array of low-loss grating waveguides and supports fully passive, two-dimensional (2D) wavelength-controlled beam steering. A fundamentally space-efficient design that folds the feed network into the aperture also enables scalable tiling of SOPAs into large apertures with a high fill-factor. We experimentally demonstrate the first SOPA, using a 1450 -- 1650 nm wavelength sweep to produce 16,500 addressable spots in a 27$\times$610 array. We also demonstrate, for the first time, far-field interference of beams from two separate OPAs on a single silicon photonic chip, as an initial step towards long-range computational imaging LIDAR based on novel active aperture synthesis schemes.
\end{abstract}

\section{Introduction}
\label{sec:intro}

\begin{figure*}[bth]
	\centering
  \includegraphics[width=1\textwidth]{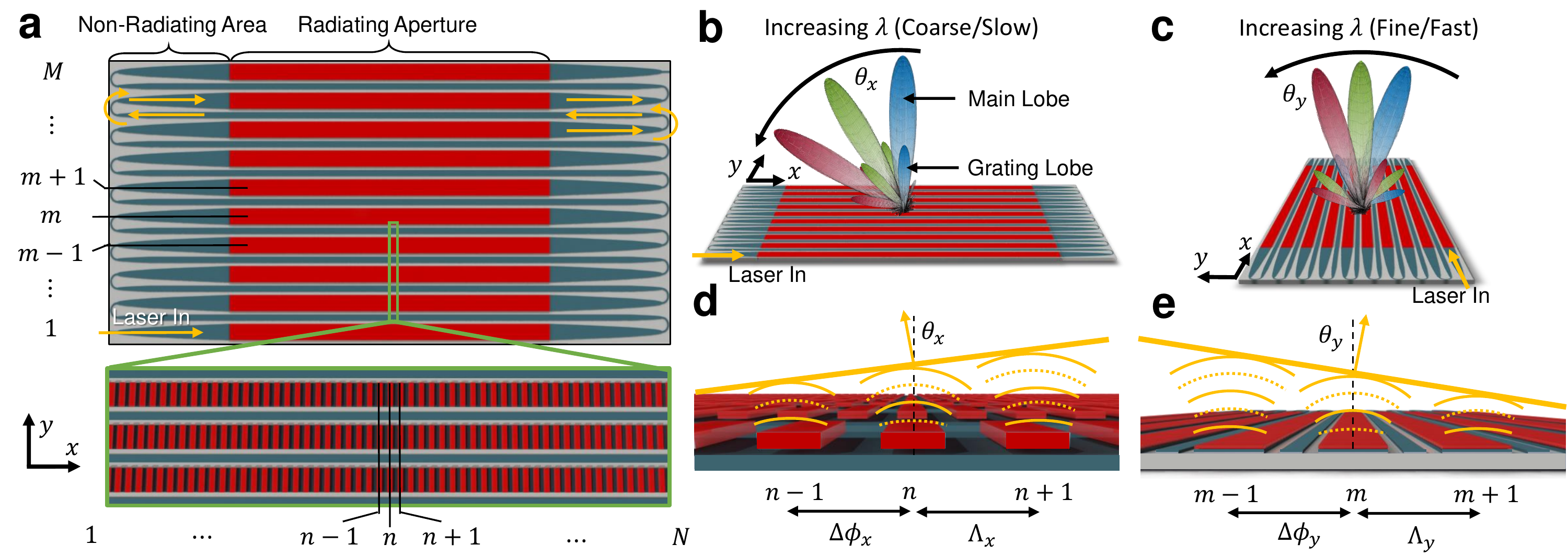}
  \caption{\textbf{Serpentine optical phased array 2D wavelength-steering.}
    \textbf{a} Schematic of SOPA tile topology. An array of $M$ rows of grating-waveguides (red) are serially connected by flybacks (blue) in a serpentine configuration. Each row has $N$ grating periods.
    \textbf{b} Coarse (slow) wavelength-steering.
    \textbf{c} Fine (fast) wavelength-steering.
    \textbf{d} For coarse steering along $\theta_x$ each grating-waveguide diffracts light to an angle determined by the wavelength-dependent tooth-to-tooth phase delay.
    \textbf{e} For fine steering along $\theta_y$ the array of gratings diffracts light to an angle determined by the wavelength-dependent row-to-row phase delay.
    }
	\label{fig:system}
\end{figure*}

\noindent Optical phased arrays (OPAs) implemented in integrated photonics \cite{van2011two,sun2013large, kwong2014chip, sayyah2015two,hulme2015fully, hutchison2016high,poulton2017large, poulton2017coherent,komljenovic2017sparse, xie2018dense,zadka2018chip,fatemi2018scalable,chung2018monolithically,phare2018silicon,komljenovic2018chip,poulton2018high,poulton2019long,poulton2019small,dostart2019vernier}  can form and electronically steer free-space optical beams to be emitted from, or received by, an on-chip aperture.  With the prospect of integration in CMOS platforms to form small size, low power, and low cost electronic-photonic systems-on-chip, integrated OPAs may become the enabling component for a new generation of photonic sensing technologies.  Furthermore, if they could scale to larger, centimeter-, reticle- or even wafer-scale apertures, silicon OPAs could make obsolete many bulk optic \cite{velodyne,terrab2015adaptive} and semi-integrated \cite{mcmanamon2009review, yoo201432, smith2017single, chen2017low,hamann2018high} approaches while enabling new, revolutionary imaging modalities \cite{wagner2019super}.

General OPA designs, capable of forming arbitrary far field patterns, may rely on 2D arrays of antenna elements \cite{sun2013large, fatemi2018scalable,chung2018monolithically,sayyah2015two} fed via independently controlled phase shifters.  They can cohere to form a single steering beam, but such approaches require an $N\times N$ array of phase shifters, each dissipating power and requiring rapid individual control, to address $N^2$ spots.  When the far field pattern of interest is one or more steered beams, a reduced control complexity is enabled by hybrid 1D phase-shifter plus 1D wavelength beam steering using an array of parallel 1D grating waveguides \cite{hutchison2016high,xie2018dense,hulme2015fully,zadka2018chip,komljenovic2017sparse,kwong2014chip,komljenovic2018chip,poulton2019long}. This approach takes advantage of the natural angular dispersion of grating couplers to steer the beam in one dimension with wavelength, thereby reducing the number of required phase shifters from approximately $N^2$ to $N$. These 1+1D OPA designs therefore use wavelength as a degree of freedom to decrease control complexity and power consumption. In addition to the remaining phase shifters, 1+1D designs demonstrated to date inherently allocate one to two orders of magnitude larger optical frequency bandwidth per spot (60 -- 100 GHz) than that needed for LIDAR ranging (1 -- 10 GHz).  This translates to a sparsely populated optical spectrum, making inefficient use of wavelength as a control parameter, and opens the door to increased noise from amplified spontaneous emission as well as a power inefficiency since the optical spectrum is contiguously generated.

Current OPA approaches have, including for some of the foregoing reasons, been limited from scaling to large (1 cm$^2$ and larger) apertures that are necessary for high performance LIDAR.  Implementing such apertures using current approaches would require, at minimum, thousands of phase-modulating elements rapidly and precisely controlled to form continuously steering beams.  The associated control complexity and power dissipation has been a significant factor in limiting achievable aperture size. Additional factors that limit aperture size are waveguide propagation loss, and the characteristic length scales of phase error accumulation across the aperture.  Construction of a large aperture as a tiled array of smaller sub-aperture tiles could take advantage of hierarchical control schemes and efficiently address some of the issues limiting aperture size.  In a suitable choice of tile and array configuration, only one phase-shifter may be needed per OPA tile to both enable steering of the tiled aperture and correct for phase errors between OPAs. However, the tiling approach can have low optical efficiency and additional beam sidelobes for apertures with low fill-factor, i.e. where substantial area is used by power feed structures and other non-emitting components.  The latter is a persistent issue for 1+1D demonstrations to date.

In this paper, we introduce the serpentine optical phased array (SOPA), a new passive and simple-to-control OPA concept designed to address these fundamental challenges and enable large-scale tiled-array apertures.  The SOPA uses a delay-accumulating, serpentine waveguide structure to wavelength-steer beams in two dimensions (2D).  The passive serpentine structure -- enabled by ultra-low-loss waveguides, tapers and single mode bends -- steers the beam along the two orthogonal angular directions via respective coarse (here, 40 GHz) and fine (1.5 GHz) frequency shifts. It nominally requires no active phase-shifters.  It also fully utilizes the optical spectrum.  The SOPA matches the ranging bandwidth to the fine beam-steering frequency increment and completely populates the utilized optical spectrum without gaps.  The 2D wavelength-steered approach thereby reduces $N^2$ degrees of freedom to just one (per beam), the optical wavelength, while using the wavelength spectrum with maximum efficiency.  It allows multiple beams to be simultaneously emitted or received by simply addressing multiple wavelengths.  The SOPA additionally uses interstitial flyback waveguides to accumulate delay across the aperture and incorporate the feed network into the OPA itself, thus allowing near unity fill-factor appropriate for tiled apertures. The increased resolution and power provided by larger LIDAR apertures can be realized by transmitting from an array of SOPA tiles that interfere on a target.  Tiling is also compatible with computational imaging techniques \cite{wagner2019super} that can correct for unknown tile phases using only the back-scattered detected signal, alleviating phase-cohering requirements. We demonstrate the SOPA concept through 2D beam steering experiments showing 16,500 addressable spots, and the tiling approach through far-field interference of beams emitted by two SOPA tiles on a single chip.  These results present a first step toward low complexity OPAs with tractable scaling to large synthetic apertures.

\section{The serpentine optical phased array}
\label{sec:tile_design}

The serpentine optical phased array (SOPA) produces two-dimensional optical beam steering by using an aperture-integrated delay-line `feed network' that in principle requires zero power and nearly zero excess footprint. It is this feature that makes the SOPA extraordinarily easy to operate and suitable to be tiled into large arrays.

An integrated OPA consists of a two-dimensional array of radiating elements with a `feed network' that distributes optical power to the elements and controls the phase of their emission for beam forming and steering. The architecture of the feed network determines the OPA's control complexity, footprint, and ultimately its scalability. Purely electronic phase control, where every radiating element is preceded by an independently-controllable phase-shifter \cite{sun2013large}, requires large numbers of phase-shifters. Frequency-based phase control uses dispersive grating couplers, delay lines, or both to map the wavelength to beam emission angle according to a frequency-dependent phase (time delay), which avoids phase-shifters entirely but `hard-wires’ the steering control to the OPA design. Most OPAs demonstrated to-date have used purely electronic steering \cite{sun2013large,fatemi2018scalable,chung2018monolithically,sayyah2015two} or replaced one dimension of steering control with wavelength by using an array of waveguide-gratings, each fed by a split and phase-shifted copy of the input signal \cite{poulton2017large,hutchison2016high,xie2018dense,hulme2015fully,zadka2018chip,raval2017unidirectional,phare2018silicon,poulton2017coherent,komljenovic2017sparse,kwong2014chip,komljenovic2018chip,poulton2018high,poulton2019long,poulton2019small}. However, the presence of electronically-controlled phase-shifters within the OPA rapidly increases the OPA’s complexity as it increases in size, making centimeter-scale apertures difficult to control.

\begin{figure*}[tb]
	\centering
	\includegraphics[width=1\textwidth]{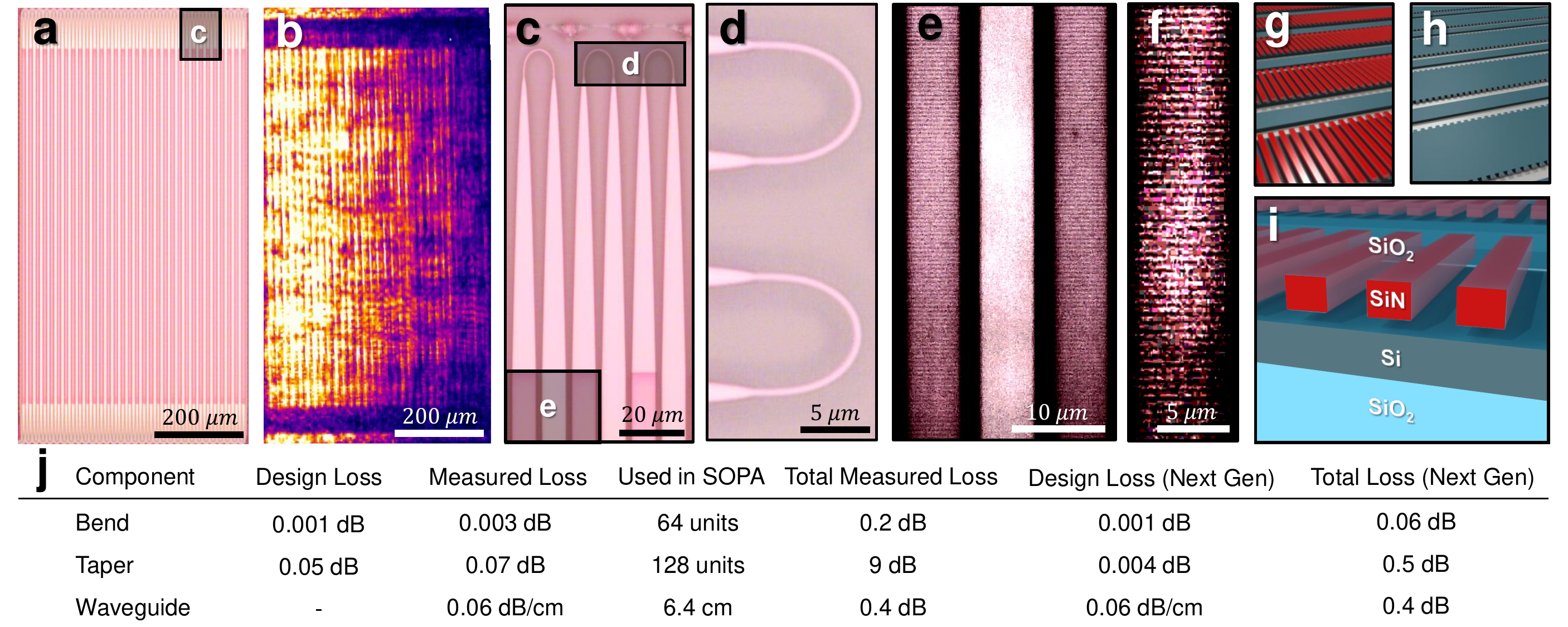}
	\caption{ 
	  \textbf{Images of fabricated SOPA emission pattern and components.}
  	\textbf{a} Optical micrograph of the fabricated SOPA where the radiating aperture is darker pink.
  	\textbf{b} Near-field IR image of the SOPA emission.
  	\textbf{c} Zoomed view of all SOPA components, centered on the adiabatic tapers, with labels for the single mode adiabatic bends and waveguides.
  	\textbf{d} Single mode adiabatic bends.
  	\textbf{e} Two grating-waveguides (left and right) and a single flyback waveguide (center).
  	\textbf{f} High magnification image of the nitride bar grating.
  	\textbf{g} Rendering of a SOPA with nitride bar grating variant.
  	\textbf{h} Rendering of a SOPA with silicon sidewall grating variant.
  	\textbf{i} Fabrication cross-section of the nitride bar grating.
  	\textbf{j} SOPA loss budget.
	}
	\label{fig:schematic}
\end{figure*}

The key to the SOPA concept is to steer with wavelength in both dimensions by using grating couplers in one dimension and a sequential folded serpentine delay line in the other. This allows the frequency of a single tunable laser to control the entire OPA, eliminating the need for phase-shifters entirely. To make the SOPA as simple and space-efficient as possible, the gratings (red) are incorporated directly into the delay line by means of a serpentine structure (blue) (Fig.~\ref{fig:system}a). Thus, unlike the initial 2D wavelength steered OPA approach which used delay lines external to the gratings \cite{van2011two}, the SOPA's delay line `feed network' incurs near zero area overhead and is independent of aperture size. We also show that the SOPA demonstrates improved performance compared to the previous 2D wavelength-steered OPA \cite{van2011two}: a $400\times$ larger aperture and $300\times$ more spots, enabling performance comparable to the state-of-the-art. This is achieved through development of ultra-low loss components in this work and optimal use of the frequency domain (each addressable spot takes up only as much bandwidth as needed for the desired ranging resolution). By removing the need for phase-shifters, and efficient use of wavelength as an easily accessible control parameter, many SOPA devices can be arrayed on a single chip to create centimeter-scale apertures which drastically outperform other OPA approaches.

The serpentine delay structure steers beams in two orthogonal dimensions by tuning the wavelength/frequency in respectively coarse and fine increments, as illustrated in Fig.~\ref{fig:system}b,c analogous to a falling raster \cite{ISI:A1990DJ52800004} demonstrated previously with dispersive reflectors in a free-space configuration \cite{ISI:000259271900033}.

The SOPA's beam steering capability is best understood in terms of the frequency-resolvability of the array, which relates the time delay across the aperture to the frequency shift required to steer by one spot. The delay accumulated along a single grating-waveguide ($\tau$) is exactly the inverse of the frequency step ($\Delta f=1/\tau$) required to steer the beam by one resolvable spot along the grating-waveguide dimension $\theta_x$ (Fig.~\ref{fig:system}d). The delay accumulated across the $M$ serpentine rows of the aperture, $T=MC\tau$ ($C$ is a constant that accounts for additional delay that may be incurred from row-to-row connecting components), therefore results in a `finer' (smaller) frequency step to steer by one resolvable spot along $\theta_y$ (Fig.~\ref{fig:system}e) than the `coarse' (large) step needed to steer along $\theta_x$. This arrangement causes the beam to steer quickly along $\theta_y$ and slowly along $\theta_x$ for a linear ramp of the optical frequency. The slow scan along $\theta_x$ combined with the periodic resetting of the steering angle along $\theta_y$ as the row-to-row phase increments by $2\pi$ results in a 2D raster scan of the FOV controlled entirely by the frequency/wavelength.

A mathematical model for 2D beam steering with frequency is obtained by considering the SOPA as a phased array. Along $x$, light is coupled out at an angle $\theta_x(f)$ through a phase matching condition:

\begin{equation}
\begin{split}
  \theta_{x}(f) 
  & = \sin^{-1}\left(\frac{c}{2\pi f}\left[\frac{\Delta\phi_x(f)}{\Lambda_x} + q\frac{ 2\pi}{\Lambda_x}\right]\right),\quad q\in\mathbb{Z}\\
  & = \sin^{-1}\left(n_{\rm eff}(f) - \frac{c}{f\Lambda_x}\right)
\end{split}
\label{eq:thetax1}
\end{equation}

\noindent where $\Lambda_x$ is the grating period, $n_{\rm eff}$ is the effective index of the waveguide mode, $q$ is the diffraction order, and $\Delta\phi_x(f)$ is the relative phase between grating periods and is given by $\Delta\phi_x(f)=2\pi f n_{\rm eff}(f) \Lambda_x/c$. We choose the grating period so that only the first diffraction order, $q=-1$, is radiating.

\begin{figure*}[tb]
	\centering
	\includegraphics[width=1\textwidth]{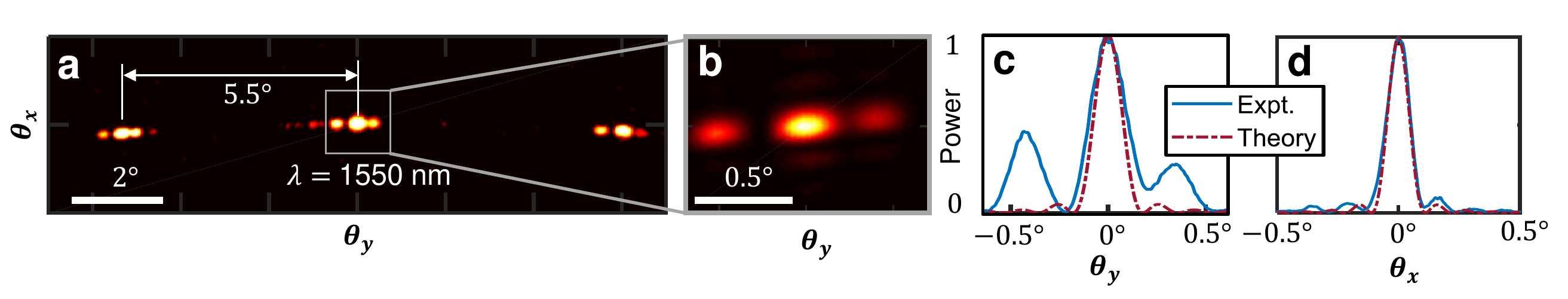}
	\caption{
	\textbf{Measurement results of single tile emission pattern.}
	\textbf{a} Example far-field emission pattern of a single tile at one wavelength showing $5.5^\circ$ grating lobe spacing.
	\textbf{b} Zoomed image of the main lobe.
	\textbf{c} Cross-section along $\theta_y$ of the main lobe with measured full-width half-max (FWHM) of $0.2^\circ$.
	\textbf{d} Cross-section along $\theta_x$ of the main lobe with measured FWHM of $0.11^\circ$.}
	\label{fig:results1}
\end{figure*}

The diffraction angle along $y$, $\theta_y(f)$, is given by:

\begin{equation}
  \theta_{y}(f)= \sin^{-1}\left(\frac{ c}{f\Lambda_y}\frac{\text{mod}_{2\pi}\left[\Delta\phi_y(f)\right]}{2\pi}\right)
\label{eq:thetay1}
\end{equation}

\noindent where $\Lambda_y$ is the row-to-row pitch, $\Delta\phi_y(f)$ is the differential phase between adjacent grating-waveguides (equal to the phase accumulated across the preceding grating-waveguide and additional connecting components), and $\text{mod}_{2\pi}[x]$ denotes the wrapped phase evaluated on the interval $(-\pi,\pi]$.

The frequency shift to steer the beam by one spot width can be found by taking the derivative of the differential phase $\Delta\phi(f)$ with respect to frequency and calculating the frequency step $\Delta f$ to create a $2\pi$ phase shift across the length of the aperture: $\Delta f_i =2\pi (\Lambda_i/L_i)( \partial \Delta \phi_i/\partial f)^{-1}$. The frequency shifts which steer the beam by one spot width along $x$ and $y$, respectively, are:

\begin{align}
  \Delta f_x &= 2\pi\left(N\frac{\partial \Delta\phi_x}{\partial f}\right)^{-1} = \frac{c}{n_{g}(f)N\Lambda_x}\\
  \Delta f_y &= 2\pi\left(M\frac{\partial \Delta\phi_y}{\partial f}\right)^{-1} = \frac{\Delta f_x}{MC}
\end{align}

\noindent where $n_{g}$ is the group index of the grating-waveguide mode, N is the number of periods along a single grating-waveguide, M is the number of grating-waveguide rows, and C is a constant that accounts for additional delay that may be incurred from row-to-row connecting components. It is clear from equations 3 and 4 that for a coarse frequency shift of $\Delta f_x$, the beam is steered by one spot width in $\theta_x$ and by M times C spot widths in $\theta_y$, during which C is the number of times $\theta_y$ scans the y-dimension FOV.


\section{Silicon photonic SOPA implementation}
\label{sec:implementation}

The first implementation of the SOPA was fabricated in a silicon-on-insulator platform, shown with its components in Fig.~\ref{fig:schematic}. An image of the full SOPA is depicted in Fig.~\ref{fig:schematic}a where the radiating aperture can be seen as the darker pink and is captured in the IR near-field image in Fig.~\ref{fig:schematic}b. The magnified image in Fig.~\ref{fig:schematic}c also contains labels for the single mode adiabatic bends (Fig.~\ref{fig:schematic}d) and grating-waveguides and flybacks (Fig.~\ref{fig:schematic}e).

The principal component of the SOPA, the grating-waveguide, is highlighted in Fig.~\ref{fig:schematic}f, with the two grating variants shown in Fig.~\ref{fig:schematic}g,h. The first weakly-scattering grating design uses an upper level of silicon nitride bars placed over the silicon waveguide as teeth (Fig.~\ref{fig:schematic}f,g), while the second uses rectangular corrugations on the sidewalls of the silicon waveguide (Fig.~\ref{fig:schematic}h). In this SOPA implementation, the gratings were designed to radiate normal to the plane of the chip at a wavelength of approximately 1300 nm (460 nm grating period).

\begin{figure*}[tb]
	\centering
	\includegraphics[width=1\textwidth]{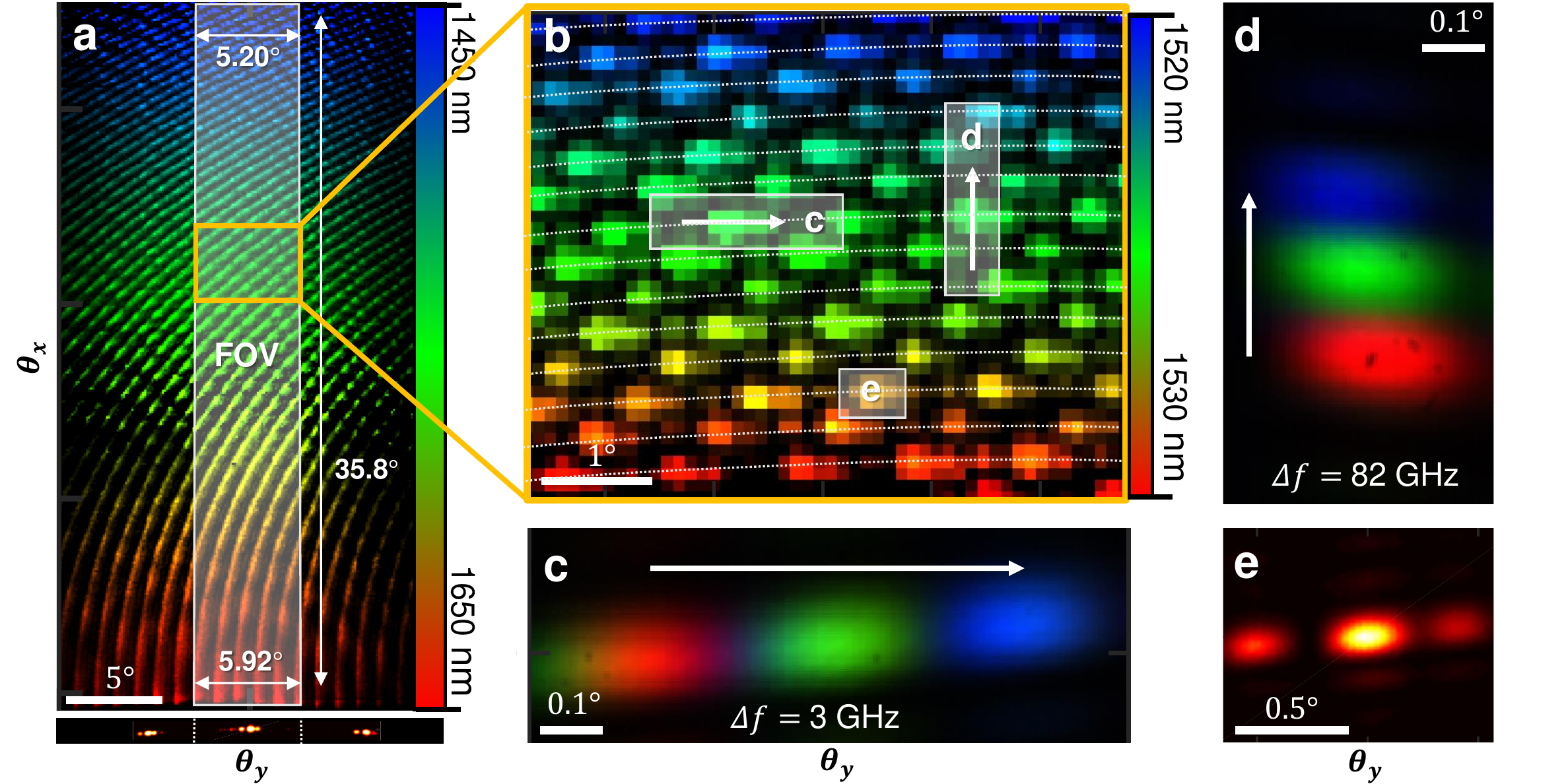}
	\caption{
	\textbf{Demonstration of 2D wavelength-steering with a SOPA.}
	\textbf{a} A far-field camera image of a $200\,$nm scan, with only 1,500 spots sampled out of 16,500. The grating lobe-limited FOV is $35.8^\circ\times5.5^\circ\,$. The under-sampling of the scan combined with group velocity dispersion causes the scan loci to appear curved. The spot pattern at 1550 nm is shown at the bottom to demonstrate the grating lobe-limited FOV.
	\textbf{b} A $5^\circ\times5.5^\circ\,$ subsection of the full scan, with only 70 spots sampled. The scan loci are depicted by the dotted lines as a guide-to-the-eye and the colors are re-coded for the narrower bandwidth.
	\textbf{c} Wavelength scanning along the fast axis with 3 non-adjacent spots spaced by $3\,$GHz.
	\textbf{d} Wavelength scanning along the slow axis with 3 non-adjacent spots spaced by $82\,$GHz.
	\textbf{e} Single-wavelength spot at 1550 nm.}
	\label{fig:scanning}
\end{figure*}

The SOPA’s serially connected grating delay-line geometry represents the key design challenge for this approach: for long optical path lengths, weak gratings and low routing losses are needed. For the results shown in this paper, the geometric optical path length is 6.4 cm with 128 tapers and 64 bends. One of the dominant sources of on-chip waveguide propagation losses is scattering due to line edge roughness. We designed wide (6.5 {\textmu}m) waveguides that exceed the single-mode width (500 nm) to mitigate this loss, similar to \cite{gevorgyan2016adc}. Adiabatic tapers, designed using the Fourier modal method \cite{Song92FMMTapers}, are used to transition from wide waveguides to single-mode width inter-row U-bends without exciting higher order modes. The inter-row U-bends are designed for ultra-low insertion loss by employing an adiabatically varying curvature \cite{cherchi2013dramatic,Zafar18TEPolarizerEulerbend} that minimizes bend loss without junction losses. The predicted and measured component losses are shown in Fig.~\ref{fig:schematic}j.

In this SOPA implementation, each SOPA tile is ${\sim}$1 mm $\times$ 0.5 mm in size with a radiating aperture of ${\sim}$0.8 mm $\times$ 0.5 mm, which results in 82\% of the footprint occupied by the radiating aperture. The length of each grating-waveguide row is 0.8 mm, and the remaining ${\sim}$0.2 mm is taken up by the adiabatic bends and tapers. The row-to-row pitch, $\Lambda_y$, is 16 {\textmu}m, and there are 32 rows of grating-waveguides and flybacks for a total tile height of 0.5 mm. The losses (visible as the exponential decay across the aperture in Fig.~\ref{fig:schematic}b) were dominated by the tapers, which have been improved in newer generations. Based on this geometry, we estimate coarse and fine frequency steps of $\Delta f_x \approx 80$ GHz and $\Delta f_y \approx 1$ GHz from Eqs.~(3,4), where a factor of $C=2.4$ is estimated based on measurements as the excess delay incurred from the flybacks, tapers, and bends.


\begin{figure*}[t]
	\centering
	\includegraphics[width=\textwidth]{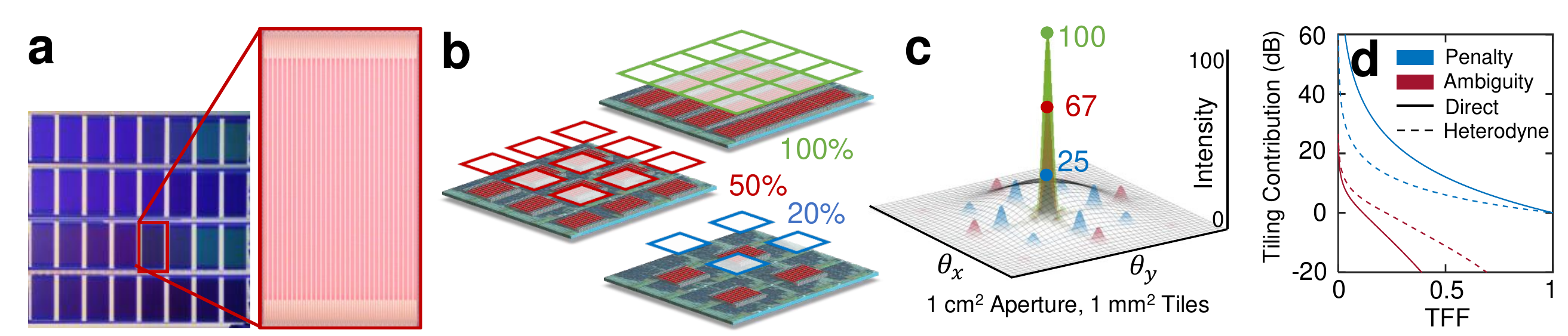}
	\caption{
	\textbf{Beam steering with tiled-apertures.}
	\textbf{a} Image of the fabricated SOPA tiled-aperture.
  \textbf{b} Illustration of tiled-apertures with varying tiling fill-factor (TFF).
  \textbf{c} Radiation patterns of tiled-apertures with TFFs from \textbf{b} relative to a single tile ($20\times$ magnified).
  \textbf{d} Relation of TFF to the contribution of tiling lobes to SNR penalty and ambiguity for both direct detection and heterodyne detection.}
	\label{fig:tiling}
\end{figure*}

\section{Single tile beam forming experiments}\label{sec:single_tile_results}

To demonstrate the low-complexity beam steering required for scaling to large apertures, we characterize the spot size, FOV, number of addressable spots, and scan rates of a single SOPA tile. Experimental results at 1550 nm of the far-field emission pattern of a single SOPA tile using the nitride bar design are shown in Fig.~\ref{fig:results1}.
We measure a full-width at half-maximum (FWHM) spot size of $\Delta\theta_x = 0.11^\circ$ by $\Delta\theta_y = 0.2^\circ$, close to the diffraction-limited size of $0.1^\circ$ by $0.16^\circ$, without any active phase tuning.
The side lobes along $\theta_y$ are due predominantly to row-to-row phase-errors accumulated along the 6.4 cm path length.
For OPAs with element pitches greater than $\lambda/2$, multiple lobes (commonly referred to as grating lobes) are emitted into the far-field and thereby limit the FOV of the OPA.
These grating lobes are seen at $\pm 5.5^\circ$ along $\theta_y$ in Fig.~\ref{fig:results1}a, limiting the FOV to $5.5^\circ$ in this direction.
Decreasing the row-to-row pitch will widen the unambiguous FOV.

Beam steering results are shown in Fig.~\ref{fig:scanning} demonstrating both a full $200$ nm wavelength sweep as well as scanning rates along both dimensions.
We measure a $\Delta f_x=41.3$ GHz frequency shift to steer to the next addressable spot along $\theta_x$ and a $\Delta f_y=1.53$ GHz frequency shift to steer to the next resolvable spot along $\theta_y$.
We measured 27 resolvable spots along $\theta_y$ (out of 32 at the diffraction limit) and 610 addressable spots along $\theta_x$ for a total of 16,470 addressable spots.


\section{Scaling up aperture size by tiling}
\label{sec:tiling}

\begin{figure*}[tb]
	\centering
	\includegraphics[width=\textwidth]{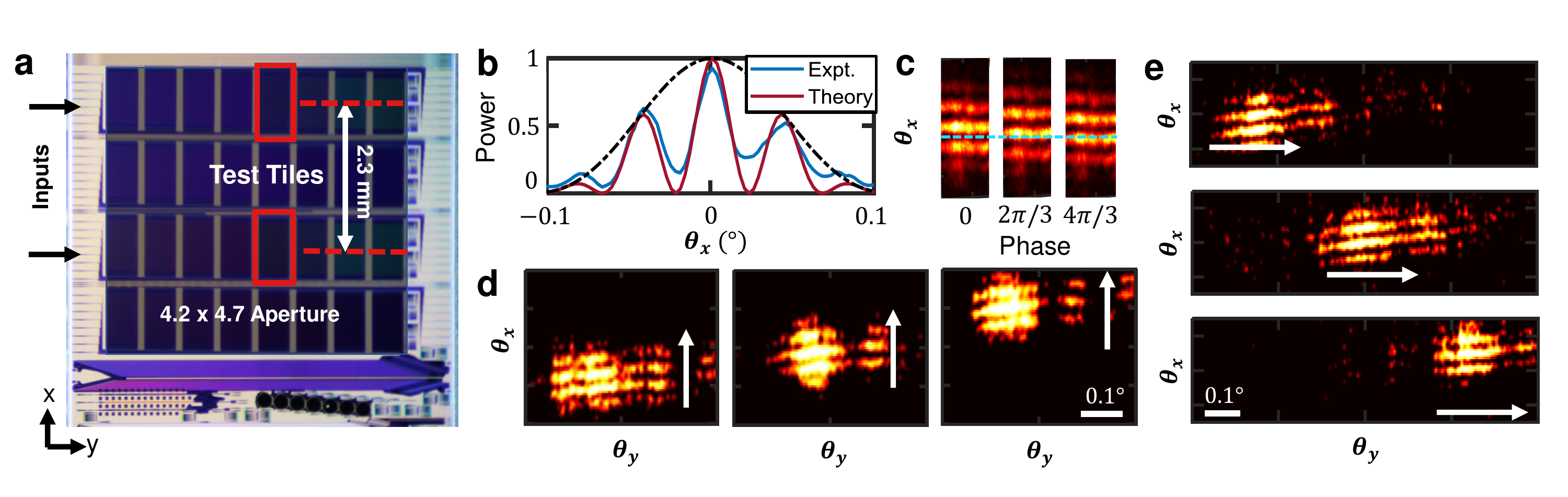}
	\caption{
	\textbf{Demonstration of tiled-aperture operation.}
	\textbf{a} Test chip showing $4.2\times4.7\,$mm\textsuperscript{2} tiled-aperture with the tiles used for the interference demonstration highlighted.
	\textbf{b} Comparison of expected and measured fringe visibility demonstrating beam balance and tile coherence.
	\textbf{c} Demonstration of relative phase-shift control (see Methods) between tiles with visible phase shift asymmetry about the dotted line.
	\textbf{d} Same-angle wavelength-steering of both tiles in the slow scanning direction ($\Delta f=82\,$GHz steps).
	\textbf{e} Same-angle wavelength-steering of both tiles in the fast scanning direction ($\Delta f=3\,$GHz steps).}
	\label{fig:results2}
\end{figure*}

In order to increase the aperture size beyond a single tile we use an array of SOPA tiles to interferometrically form a composite aperture \cite{wagner2019super}. The tiling approach provides a means to create a larger effective aperture than is feasible with a single contiguous OPA tile with a scheme which can be extended to arbitrarily large aperture sizes. For example, our fabricated test aperture shown in Fig.~\ref{fig:tiling}a is approximately 20 mm$^2$, much larger than can be achieved with a single SOPA tile (0.5 mm$^2$) which could be limited in size by waveguide loss, phase error accumulation, and, for LIDAR applications, ranging bandwidth.

The tiling approach is conceptually simple: an array of identical OPAs is fed in parallel by a distribution network (e.g. splitter tree) with a single input laser and each OPA is preceded by a single phase-shifter to facilitate array-level beam steering. All OPAs, emitting simultaneously, beam steer to the same far-field spot to create overlapping and coherently interfering spots on the target. This beam alignment of multiple OPAs is automatic for 2D wavelength-steered OPAs, such as the SOPA, when driven by a common laser; phase-shifter steered OPAs require additional control to ensure beam alignment.
When all tiles emit with identical phase a smaller `array' spot is formed in the center of the overlapping `tile' spots. Linear phase ramps applied (via the array-level phase-shifters) across the array of tiles along $x$ and $y$ will steer the array spot in 2D within the bounds of the tile spot, enabling imaging within the tile spot.
The array of tiles thereby creates a larger effective aperture composed of individual OPAs, the tiled-aperture, achieving correspondingly higher overall tiled-aperture resolution.

One potential downside of the tiling approach is the creation of additional radiation lobes (`tiling lobes'), which are directly analogous to the grating lobes produced by the gaps in element spacing in OPAs \cite{sun2013large,fatemi2018scalable}.
These tiling lobes affect system performance by diverting transmitted (or received) power away from the main lobe which decreases the signal-to-noise ratio (SNR) and introduces ambiguity to a LIDAR or imaging system. These effects are minimized when the separation between radiating elements of adjacent tiles is minimized, which can be quantified by a `tiling fill-factor' (TFF) metric we introduce as
\begin{equation}
  \text{TFF} \equiv \frac{ \text{area of emitting sub-aperture} }{ \text{total tile area} }
\label{eq:tff}
\end{equation}

\noindent The TFF should not be confused with the fill-factor of the OPA tile's radiating aperture, a separate metric which captures the density of radiating elements within a single OPA tile and correspondingly correlates with the power radiated into grating lobes. The product of the TFF and the OPA tile's fill-factor captures the combined effects of both tiling and grating lobes.

The TFF directly correlates with the aforementioned effects of decreased SNR and increased ambiguity, as shown in Fig.~\ref{fig:tiling}b-d for a LIDAR system using identical tiled-apertures for both the transmitter and receiver. For an OPA composed of rectangular apertures and direct detection the `tiling penalty' to SNR scales proportionally to $1/\text{TFF}^4$. Defining the ambiguity of the measurement as the ratio of erroneous signal (due to tiling lobes) to actual signal (due to the main lobe), the same system has a `tiling ambiguity' proportional to $1/\text{TFF}$ for low TFF. It is therefore essential to use OPA designs with high TFF when using the tiled-aperture approach.


\section{Two-tile aperture synthesis experiments}
\label{sec:dual_tile_results}

We demonstrate the creation of a larger effective aperture using the tiled-aperture approach by transmitting simultaneously from two SOPA tiles (Fig.~\ref{fig:results2}). These results constitute, to the best of our knowledge, the first demonstration of tiled-aperture optical beam steering.

The tiles are located on the same chip with center-to-center separation of 2.3 mm, as shown in Fig.~\ref{fig:results2}a.
The measured far-field fringe pattern is shown in Fig.~\ref{fig:results2}b and demonstrates phase uniformity within the far-field spot, an essential ingredient for aperture synthesis.
Fig.~\ref{fig:results2}c depicts shifting of the fringe pattern as the input phase to one of the tiles is changed, demonstrating the phase control necessary for both diffraction-limited tiled-aperture beam forming and `super-resolved' structured illumination imaging within a single far-field spot \cite{wagner2019super}.
Although we use off-chip phase control for these initial experiments, future designs will include active integrated phase-shifters for this purpose.
An additional requirement for tiled-aperture operation is that every tile steers to the same far-field spot, which we demonstrate with our two tiles in Fig.~\ref{fig:results2}d,e.
As the beams steer, the fringe visibility remains high indicating that the beams remain coherent and power-balanced as the wavelength is tuned.


\begin{table*}[tb]
	\centering
  \begin{tabular*}{\textwidth}{l @{\extracolsep{\fill}} ccccccccc}
  
    & \cite{sun2013large} & \cite{fatemi2018scalable} & \cite{kwong2014chip} & \cite{hulme2015fully} & \cite{hutchison2016high} & \cite{poulton2019long} & \cite{xie2018dense} & \cite{van2011two} & This Work \\
    \hline\\
      
    Steering Method & \multicolumn{2}{c}{\textit{2D Phase-Shifter}} & \multicolumn{5}{c}{\textit{1D Wavelength, 1D Phase-Shifter}} & \multicolumn{2}{c}{\textit{2D Wavelength}} \\
    \cline{2-3}\cline{4-8}\cline{9-10}\\
    & \multicolumn{9}{c}{\textbf{Single Aperture Performance}}\\
    Phase-Shifters & 64 & 128 & 16 & 32 & 128 & 512 & 32 & 0 & 0 \\
    Radiating Area (mm\textsuperscript{2}) & 0.0052 & 0.023 & 0.0048 & 0.032 & 0.49* & 10 & 1.3 & 0.001* & 0.4 \\
    Spots & 64 & 400 & 500 & 140 & 60,000 & 190,000* & 45,000 & 50 & 16,500 \\
    FOV ($^\circ$) & 6.0 x 6.0 & 16 x 16 & 20 x 15 & 23 x 3.6 & 80 x 17 & 15 x 56 & 28 x 22 & 15 x 50 & 36 x 5.5 \\
    \\
    & \multicolumn{9}{c}{\textbf{$1\,$cm\textsuperscript{2} Tiled-Aperture Extrapolated Performance}}\\
    \hline\\
    Tiling Fill Factor & 0.11$^\dagger$ & 0.18$^\dagger$ & 0.003* & 0.005* & 0.083* & 0.70* & 0.31* & 0.2* & {\textbf{0.82}} \\
    Phase-Shifters & 1,300,000 & 570,000 & 1,100* & 500* & 2,100* & 3,600* & 790* & 20,000* & {\textbf{200}} \\
  \end{tabular*}
  
      

  \caption{Performance comparison of demonstrated 2D integrated beam steering apertures.\\ {\small*Value estimated from figures or other stated parameters. $^\dagger$2D phase-shifter tiles have TFF identical to the tile internal fill-factor.}}
	\label{tab:comparison}
\end{table*}

\section{Discussion}
\label{sec:disc}

To demonstrate the SOPA's scalability to larger apertures via tiling, its performance is compared to selected results from the literature in Table~\ref{tab:comparison}, both as a standalone aperture and when multiple tiles are combined to form a 1 cm\textsuperscript{2} sized tiled-aperture as needed for moderate range ($\sim100$ m) LIDAR. The TFF of $82\%$ in this initial demonstration and use of 2D wavelength-steering at the single tile level lead to high-density tiling and far fewer phase-shifters than competing approaches at the tiled-aperture level. A 1 cm$^2$ SOPA tiled-aperture using the initial SOPA tile design in this paper requires only 200 phase-shifters, one per tile.
By comparison, approaches using phase-shifter steering generally require large numbers of phase-shifters ($>$1,000) or have degraded performance due to low TFF ($<10\%$).
Only two recent designs \cite{poulton2019long,xie2018dense} have both a high TFF and manageable number of phase-shifters at the centimeter aperture scale.


As a standalone aperture, the SOPA excels in its minimal electronic complexity, while achieving aperture size, number of spots, and FOV comparable to the state-of-the-art.
Several improvements have been made to the SOPA design for increased performance in future generations.
Interleaved SOPAs, each addressing a half-space, allows doubling of the slow axis angular scan range to nearly 80$^\circ$.
The FOV along the fast dimension has been widened to 18$^\circ$ by reducing the row-to-row pitch down to 5 {\textmu}m and optimizing for low waveguide, taper, and bending loss.
Designs with shortened tapers achieved upwards of 92\% TFF.
Improvements to spot quality, which is currently not diffraction-limited due to fabrication non-uniformity, can be realized with only 2 to 4 thermal phase tuners directly integrated into the SOPA without reducing TFF.
More efficient SOPA delay accumulation ($C\to1$) will reduce the phase errors and may be sufficient to avoid phase error compensation entirely.
Apodized and multi-level grating designs \cite{wade201575} will enable unidirectional emission, maximum power extraction, and improved beam profiles.
Other improvements include spatial Vernier topologies for grating lobe suppression in bidirectional LIDAR \cite{dostart2019vernier} and a Fourier-basis self-calibrated tiled-aperture imaging approach \cite{wagner2019super}.

In this paper, we demonstrated the SOPA silicon photonic wavelength controlled 2D beam steering tile, an OPA designed from the outset for low-complexity, large-area, tiled-apertures.
Our delay-accumulating serpentine waveguide design allows for compact, wavelength-steered tiles which pack efficiently into an array and do not require active control.
We demonstrated a SOPA tile with nearly diffraction-limited performance, 16,500 addressable spots, wide FOV, and record tiling density, as well as interferometric beam combination with two-tile simultaneous beam steering emulating aperture synthesis imaging from a large array of SOPA tiles.
We believe that the SOPA design is a promising solution for achieving easily controllable, large, 2D beam steering apertures demanded by applications such as long range integrated photonic LIDAR.

\vspace{8pt}

\noindent \textbf{\large Funding.}
U.S. Government; National Defense Science and Engineering Graduate Fellowship Program (NDSEG) \\(GS00Q14OADS139); National Science Foundation (NSF) (1144083); Packard Fellowship for Science \& Engineering (2012-38222).

\vspace{8pt}

\noindent \textbf{\large Acknowledgments.}
Chip layout was carried out using an academic license of Luceda Photonics IPKISS.


\bibliographystyle{ieeetr}
\bibliography{biblio.bib}

\end{document}


\maketitle

\begin{abstract} 
This document provides supplementary information to ``Serpentine optical phased arrays for scalable integrated photonic LIDAR beam steering.''
The supplementary material includes measurement details for SOPA components and OPA performance. Additionally this document discusses the limitations on OPA performance imposed by grating and tiling lobes in the form of SNR and ambiguity as well as lobe suppression techniques. The definition of fill-factor used in the main text is derived in full here. Potential detection techniques for tiled array transceivers are also presented.
\end{abstract}

\section{Component Loss Measurements}

The SOPA consists of four main optical components - waveguides, gratings, tapers and bends - that were serially connected together, forming a long optical path length of 6.4 cm in the design demonstrated here.
Because the optical path is so long (${\sim40,000}$ times the wavelength), the components needed to be designed for very low loss such that the light was able to traverse the entire SOPA path and sufficiently illuminate the aperture.

\begin{figure}[t!]
	\centering
	\includegraphics[width=1\textwidth]{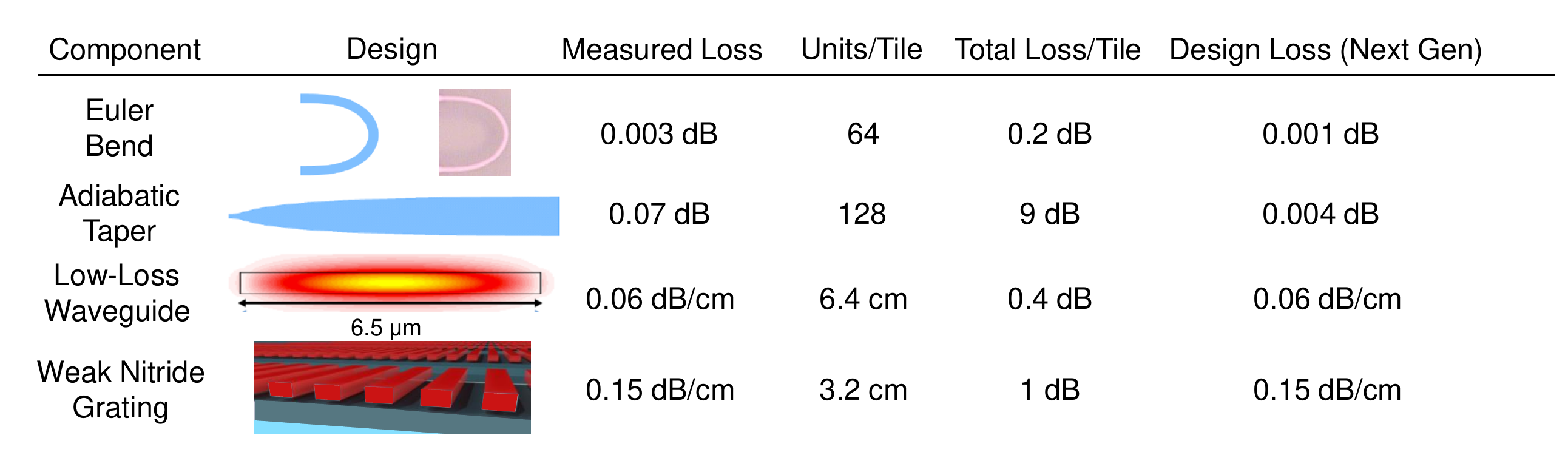}
	\caption{\textbf{Component Loss Measurements.}}
	\label{fig:comp_loss}
\end{figure}

The four components and their measured losses are summarized in Fig. \ref{fig:comp_loss}.
For low-loss routing in the grating-waveguides and the flybacks, the waveguides were made to be 6.5 {\textmu}m wide so that the fundamental mode of $1/e^2$ width 3 {\textmu}m was highly confined to the center of the waveguide.
By confining the mode to the center of the wide waveguide, the mode had less interaction with the sidewalls and less surface roughness-induced scattering loss \cite{gevorgyan2016adc}.
We measured propagation loss of approximately 0.06 dB/cm for 2 {\textmu}m wide waveguides, which to our knowledge is a record for the AIM process. Propagation loss for 6.5 {\textmu}m wide waveguides was too low to accurately measure.
Low-loss $180^\circ$ bends with adiabatically changing curvature \cite{cherchi2013dramatic,Zafar18TEPolarizerEulerbend} are designed to connect rows together.
The waveguide width for the bends was chosen to be sufficiently small to only support the fundamental mode: 0.5 {\textmu}m. The input to output pitch of the bends was 8 {\textmu}m, setting the row-to-row pitch as 16 {\textmu}m.
We measured an insertion loss of 0.003 dB per bend.
Adiabatic tapers were designed to connect the 0.5 {\textmu}m wide bends to the 6.5 {\textmu}m wide waveguides.
We measured a taper insertion loss of 0.07 dB per taper in our first implementation, resulting in the tapers contributing the large majority of loss across the full SOPA aperture (9 dB).

For low-loss gratings, various grating teeth were used. Sidewall perturbations ranging from 10 nm to 200 nm were used as a silicon-only grating design with grating strength up to approximately 1 dB/cm. A sidewall perturbation of 100 nm was used for the two tile results in the main text. Other grating designs used bars in one of the two nitride layers as grating teeth. The nitride layer which was closer to the silicon waveguide had a grating strength upwards of 30 dB/cm which radiated all of the light in the first several rows. The nitride layer which was further from the waveguide had a lower perturbation and has an estimated grating strength of 0.15 dB/cm, allowing approximately uniform emission from the SOPA in the absence of component losses. This second, weaker nitride grating was used for the single tile results. Future designs will be optimized to achieve $0.5-1$ dB/cm unidirectional grating strength by utilizing multiple layers for the grating teeth.

\section{Grating Lobes}
\label{sec:grating_lobes}

For element pitches greater than $\lambda/2$, which cannot be achieved in the row-to-row spacing of the SOPA, multiple lobes are emitted into the far-field, commonly referred to as grating lobes. All 2D integrated photonic OPAs demonstrated to-date have been unable to meet this spacing requirement, and therefore grating lobes are common to all demonstrations so far. 
The grating lobe spacing along the fast axis of the SOPA limits the FOV and is related to the row-to-row pitch by $\Delta\theta=\sin^{-1}(\lambda/\Lambda_y)$.
While the grating lobe spacing is set by the row-to-row pitch, the distribution of optical power emitted into the grating lobes is determined by the element pattern and spacing of the gratings.
For a uniform grating-waveguide of width $w$ the proportion of radiated power which is located in grating lobes is approximately $P_\text{gl} = P_\text{rad}(1-w/\Lambda_y)$. Notably this ratio is the fill-factor FF$_T$ of the of the OPA, see Sec.~\ref{subsec:fill_fac} for additional discussion.
To minimize the power radiated to grating lobes we therefore require to maximize the ratio of the grating-waveguide width to the grating-waveguide spacing.

The SOPA grating lobes, due to the relatively large element spacing of $16\,${\textmu}m, significantly limit the FOV to $5.5^\circ$ along $\theta_y$. With a grating width of $6.5$ {\textmu}m, the SOPA fill-factor is $40.6\%$ corresponding to just under $60\%$ of the radiated power being emitted into grating lobes. Grating lobes can be suppressed in order to avoid potential ambiguity and to recover the full FOV using a variety of approaches, discussed further in Sec.~\ref{sec:grating_lobe_suppress}.

Without a suppression approach, grating lobes introduce measurement ambiguity directly analogous to the ambiguity associated with tiling lobes, discussed further in Sec.~\ref{subsec:ambiguity}. There is also power loss due to grating lobes which will decrease SNR in a LIDAR system, again analogous to the power loss due to tiling lobes discussed further in Sec.~\ref{subsec:snr}. In particular, we can perform the exact same derivations from Secs.~\ref{subsec:snr}-\ref{subsec:ambiguity} simply substituting the fill-factor of the tile (FF$_T$ in Sec.~\ref{subsec:fill_fac}) for TFF and arrive at identical conclusions with substituted fill-factor.

\begin{figure}[t!]
	\centering
	\includegraphics[width=.6\textwidth]{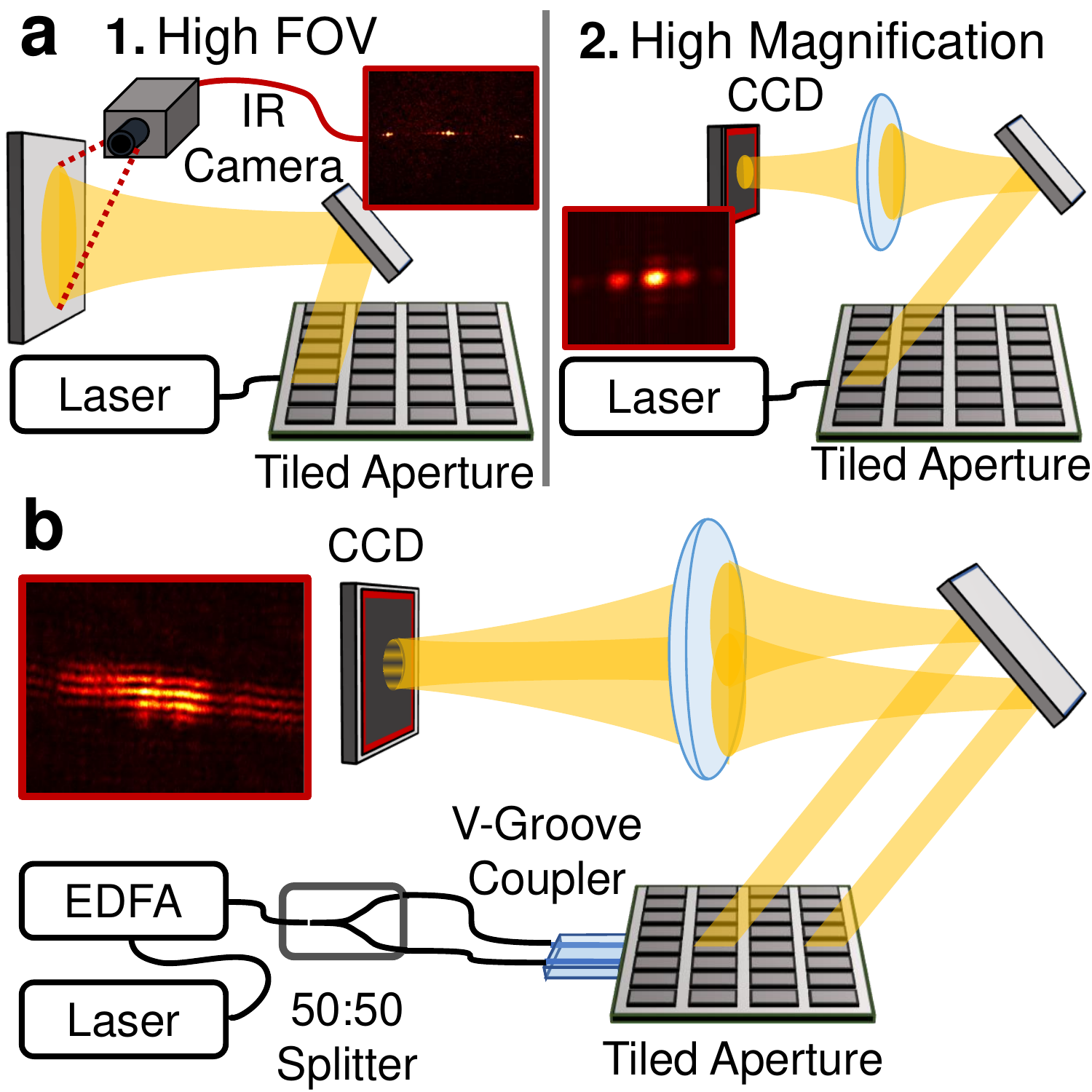}
	\caption{\textbf{Experimental setups for single and two tile measurements.} \textbf{a} Single tile measurement setup; high FOV measurements (1), such as scanning and grating lobe measurements, utilize an IR camera imaging the spot on a far-field plane. Light is coupled into the chip using a stripped, standard $8\,${\textmu}m core SMF abutted to an edge coupler. High resolution measurements (2) use a Fourier imaging configuration where the spot is incident on a CCD by placing the lens one focal length away from the CCD. \textbf{b} Two tile measurement setup, using the Fourier imaging configuration with EDFA amplification and a 50:50 splitter, coupled into the chip via a fiber v-groove.}
	\label{fig:setup}
\end{figure}

\section{Methods}
\label{sec:methods}

Our devices were fabricated in the AIM Photonics multi project wafer (MPW) process (run Feb. 2018), which was chosen for its low waveguide loss and flexibility to use multiple independently pattern-able device layers \cite{aimpdk}.

We tested the performance of individual tiles and pairs of tiles by imaging the far-field distribution with the setups shown in Fig.~\ref{fig:setup}. We used two different optical setups to image the far-field: a matte planar surface at approximately the Rayleigh range, and a Fourier transform lens (IR, 100 mm focal length, 0.45 NA) focusing onto the InGaAs SWIR detector (Photon Focus MV3-D640I-MO1-144-G2). In the first case, the surface was placed beyond the Rayleigh range and imaged using the InGaAs detector with a camera objective (Fig.~\ref{fig:setup}a1). In the second case, the lens was placed between the tile and the focal plane, with one focal length on each side (Fig.~\ref{fig:setup}a2,b). The first setup allows for a larger FOV in order to capture scanning but is complicated by trigonometric projection factors resulting in curved beam steering loci, while the second gives a higher resolution image of the spot and spot overlap, as well as more rectilinear scanning.

In both cases, the tile is placed on a positioning stage and light is coupled in via an optical fiber into AIM PDK component library standard fiber edge couplers \cite{aimpdk}. A mirror is used to reflect the beam into the plane of the optical table and onto the far-field surface or into the lens. A tunable laser (Keysight 8164B Lightwave Measurement System with 81608A Laser Module) is used to sweep the wavelength and thereby the angle in 2D; for the two-tile case, the laser passes through an optical amplifier (Amonics AEDFA-HP) and 50:50 splitter before entering the chip through a fiber v-groove array (Meisu 16 channel PM fiber array, 127 {\textmu}m pitch, Corning PM15-U25D fiber). In this initial demonstration, relative phase shift between the tiles is introduced by straining one of the fibers following the splitter.

All presented images have had a measured background image subtracted from the raw data to remove fixed extraneous scattering and camera hot pixels. For the 200nm sweep figure, we computationally equalize the power in the main lobe to correct for variations in laser power and radiation efficiency across the wavelength range. The scan image is generated by taking $1,500$ individual images at a constant $17\,$GHz frequency interval (and subtracting the background), pseudo-color-encoding each image according to the wavelength, and creating a composite image by adding all the individual images.

\section{Spot Measurements}
\label{sec:spot_measurements}

In order to determine spot size, the number of addressable spots attainable by the SOPA implementation, the frequency steps required for scanning along the fast and slow directions, the FOV, and dual tile operation several versions of experimental setups were used, discussed in Sec.~\ref{sec:methods}. For both single- and dual-tile measurements where a detailed image of the spot is needed, a long focal length lens is used to measure the far-field by placing a CCD in the Fourier plane of the lens. For measurements where a large angular space needed to be viewed simultaneously, i.e. to measure the full SOPA FOV of a 200 nm scan or individual fast and slow scans, either a short focal length lens with low F number is used or a plane is placed at approximately the Rayleigh range ($z_0=w^2/\lambda$) and imaged with an IR camera. For these measurements the plane was placed such that the beam travels approximately 51 cm to the plane, where the distance to the plane was limited by the space available in the lab. For the figures presented in this work $\theta_x=0$ corresponds to a beam which is normal to the measurement plane. For a tile with aperture 0.8 mm $\times$0.5 mm the Rayleigh range is about 43 cm along $x$ and 17 cm along $y$, so this plane is not truly in the far-field which requires $\sim10z_0$.

\begin{figure}[t!]
	\centering
	\includegraphics[width=.8\textwidth]{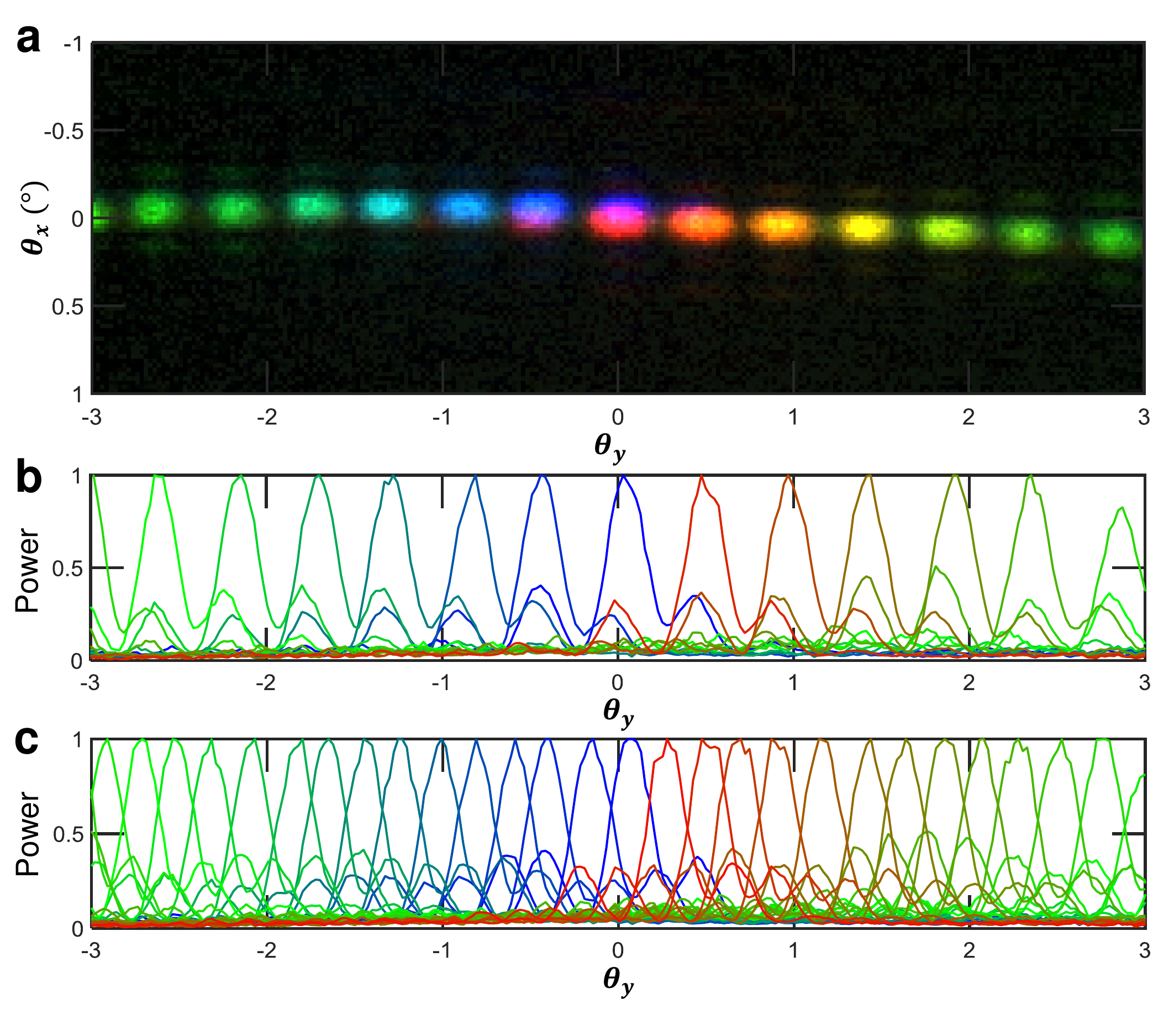}
	\caption{\textbf{Fast scan measurements.} \textbf{a} False color coded fast scan with 14 spots at 1550 nm with a 41.3 GHz sweep where the 14\textsuperscript{th} spot is the next slow scan spot. \textbf{b} Projection along $\theta_x$ to acquire spot cross-sections for 13 of the spots in part a with normalized power. \textbf{c} The same cross-section with 27 spots, spaced at 1.53 GHz intervals, demonstrating our claim of 27 resolvable spots along the fast axis.}
	\label{fig:fast_scan}
\end{figure}

We performed slow scan measurements using the lens-less setup and fast scan measurements with the Fourier lens setup to determine the number of addressable spots attainable by the SOPA with a 200 nm wavelength sweep. The fast scan results can be found in Fig.~\ref{fig:fast_scan} for a fast scan at 1550 nm. The SOPA operates over an even wider wavelength range, with normal emission at $\lambda_N=1300$ nm, but our laser is limited to a 200 nm wavelength scan range from $1450-1650$ nm. Notably, in Fig.~\ref{fig:fast_scan}a where every other resolvable spot is shown it can be seen that a fast scan across 41.3 GHz results in a spot which is aligned along the fast axis with the initial spot but shifted along the slow axis to the next addressable spot. This demonstrates the fast/slow raster scanning pattern desired. Taking a projection of each spot image along $\theta_x$ allows for showing the spot cross-section with normalized spot powers in Fig.~\ref{fig:fast_scan}b, and with every resolvable spot in Fig.~\ref{fig:fast_scan}c. This 27 spot fast scan (where the 28\textsuperscript{th} spot is the next slow scan spot) demonstrates the FWHM resolvability of the 27 spots along the fast direction. A key point is that the spot width and grating lobe spacing scale together at different wavelengths, and the number of resolvable spots is invariant across the 200 nm sweep. For an ideal realization there will be as many resolvable spots as grating-waveguide rows (here, 32).

\begin{figure}[t!]
	\centering
	\includegraphics[width=1\textwidth]{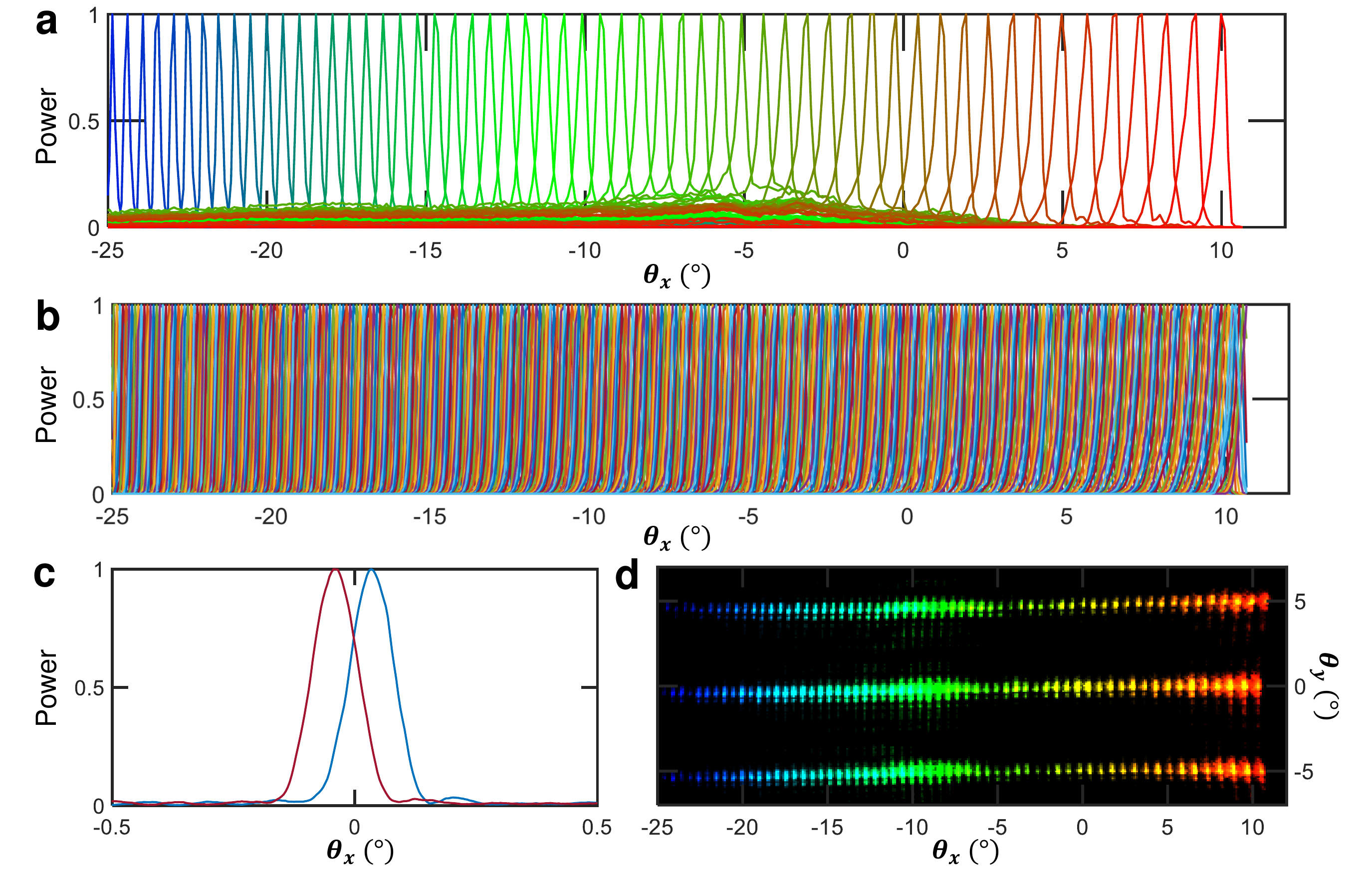}
	\caption{\textbf{Slow scan measurements.} \textbf{a} Projection along $\theta_y$ to acquire spot cross-sections for every 10\textsuperscript{th} spot along a slow scan (62 of 610 spots) with normalized spot power. The spot widths are exaggerated here due to detector saturation and a measurement plane not in the far-field. \textbf{b} The same cross-section with all 610 spots, spaced at intervals ranging from 41.7 GHz to 40.8 GHz from 1450 nm to 1650 nm. \textbf{c} Two adjacent, power-normalized slow scan spots at 1550 nm measured in the far-field (Fourier plane) demonstrating the resolvability of the addressable spots. \textbf{d} False color coded slow scan image with every 10\textsuperscript{th} spot.}
	\label{fig:slow_scan}
\end{figure}

Slow scan measurements are shown in Fig.~\ref{fig:slow_scan} across the full 200 nm sweep. A slow scan where only every 10\textsuperscript{th} spot (62 spots total) is sampled is shown in cross-sectional view, where the spot powers are normalized, in Fig.~\ref{fig:slow_scan}a and 2D image of the scan is shown in Fig.~\ref{fig:slow_scan}d. The full slow scan, with all 610 spots, is shown in cross-sectional view in Fig.~\ref{fig:slow_scan}b. Notably the frequency spacing required to move by one slow scan spot varies across the 200 nm wavelength range due to group velocity dispersion and the wavelength increment is inherently nonlinear across such a wide wavelength range. We measured 40.8 GHz frequency steps at 1450 nm, 41.3 GHz frequency steps at 1550 nm, and 41.7 GHz frequency steps at 1650 nm to obtain the approximately `straight' slow scan pattern depicted in Fig.~\ref{fig:slow_scan}d. Because these frequency spacings do not vary linearly, both 1\textsuperscript{st} and 2\textsuperscript{nd} order group velocity dispersion terms must be accounted for in this large wavelength range.

To demonstrate the number of addressable slow scan spots, two slow scan spot cross-sections are shown in Fig.~\ref{fig:slow_scan}c using the Fourier plane setup. The two spots cross at approximately 70\% of peak power, and are nearly Sparrow resolvable (crossing at 50\% of peak power). The slow scan spot spacing is controlled by the relative amount of delay in the flybacks as compared to the grating waveguides, and later designs will reduce this ratio to increase spot resolvability along the slow axis such that all addressable spots are fully resolvable.

\section{Tiling Lobes}
\label{sec:tiling_lobes}

When the separation of phased array elements (in this case, tiles) is larger than $\lambda/2$, the array will produce radiation lobes: additional beams that radiate to undesired angles, thereby reducing power in the desired beam and increasing signal ambiguity \cite{johnson1984antenna}. These `tiling lobes' are directly analogous to the grating lobes formed by element spacing larger than $\lambda/2$ within a single OPA tile.
We quantify the tiling density using the tiling fill-factor (TFF), defined in the main text and discussed in detail in Sec.~\ref{subsec:fill_fac}.

For a rectangular grid of rectangular apertures, it can be seen that there are $N \approx 1/\text{TFF}$ beams radiated within the sinc envelope of a single aperture, and the main lobe carries $1/N$ of the radiated power (see also \eqref{eq:field_pattern} in Sec.~\ref{subsec:snr}). This can also be seen from the directivity of the tiled array into the hemisphere \cite{johnson1984antenna} which is approximately $D=(\text{TFF})2\pi z^2/A$ for a spot at a distance $z$ with area $A$. Since the TFF is a metric which captures the number of tiling lobes, and correspondingly power loss to tiling lobes, it provides a useful metric of array performance.

Two main issues arise due to tiling lobes: signal power loss and ambiguous measurements due to the presence of erroneous return signals from tiling lobes. The first effect is a power loss penalty (decreased SNR) due to tiling lobes and discussed in Sec.~\ref{subsec:snr}. The second effect is detrimental in that erroneous signal (due to tiling lobes) can hide the desired signal (due to the main lobe), obfuscating the measurement and discussed in Sec.~\ref{subsec:ambiguity}.

\subsection{Fill-Factor Definitions}
\label{subsec:fill_fac}

In this section we wish to define the fill-factors introduced in the text in a rigorous manner. We additionally desire to design a fill-factor expression which applies equally well to a single OPA tile, or a tiled-array of OPAs. For a single OPA tile we will refer to an example design which is composed of $M$ grating rows of length $L$, width $w_T$, and periodicity $\Lambda_T> w_T$. For a tiled aperture we will use an array of identical tiles in a rectangular pattern with $N_x\times N_y$ individual tiles at periods of $D_x$, $D_y$ with aperture widths $W_x$, $W_y$ with each tile emitting the same field pattern $U_{T}$.

For an arbitrary OPA design, we wish to define a fill-factor which defines the ratio of the emitting area (effective aperture area) of the OPA to the total area taken up by the OPA. We can define this fill-factor as

\begin{equation}
    \text{FF} = \frac{A_{eff}}{A_{tot}}.
\end{equation}

\noindent In order to clarify this definition, we require definitions of both the area terms.

We begin by finding a mathematical definition of effective radiating (aperture) area $A_{eff}$, the emitting area of the OPA, to use for the numerator of the fill-factor. Henceforth we refer to this area as simply the `effective area' for short. For OPAs which emit at a uniform amplitude, such as apertures which can be represented in terms of rect and circ functions, this effective area can be calculated by inspection by summing up the area of these window functions. For the single OPA tile example the effective area is $A_{T,eff} = Mw_TL$, and for the tiled-array the effective area is $A_{A,eff} = N_xN_yA_{T,eff}$. For non-uniform emission, however, the effective area is not clear upon inspection and we therefore require a mathematical expression. 

We define the effective area of a non-uniform field to align with the uniform case. For a uniform field with unknown, constant amplitude $\bar{U}$ and unknown area $A_{eff}$ we can find the effective area by considering two integrals: the field power $\iint |U(\vec{r})|^2 d\vec{r} = \bar{U}^2A_{eff}$ and the integral over square intensity $\iint |U(\vec{r})|^4 d\vec{r} = \bar{U}^4A_{eff}$. While the first integral is chosen as it is directly proportional to the optical power, the second integral we choose out of convenience as this term will appear in our calculations of detected signal power. We can then find the effective area as the square of the first integral divided by the second $A_{eff} = (\bar{U}^2A_{eff})^2/\bar{U}^4A_{eff}$. The full integral from is then

\begin{equation}
\label{eq:eff_area}
    A_{eff} = \frac{\left(\iint |U(\vec{r})|^2 d\vec{r}\right)^2}{\iint |U(\vec{r})|^4 d\vec{r}}.
\end{equation}

If $U$ is normalized such that $\iint |U|^2 d\vec{r}=1$, then the effective area is simply $A_{eff}=1/\iint |U|^4 d\vec{r}$. For a non-uniform field, this definition finds the area of an equivalent, uniform aperture while maintaining the power of the field. As an example, this definition of effective area gives $A_{eff}=\pi w^2$ for a circular Gaussian of the form $U = U_0\exp[-r^2/(w^2)]$.

With a definition of the effective area, we now require to similarly define the total area of the OPA, the total footprint on-chip required for routing, modulation, and the aperture (effective area) itself. However, we will find that there is no obvious way to explicitly, mathematically define the total area for the generic case of an array of non-uniform tiles. We are able to define the total area in general terms, and for a specific OPA emitted field the total area can be found.

There are two relevant choices for defining the total area of an OPA: the minimum-area shape which encloses the radiating aperture, and the minimum-area shape which encloses both the radiating aperture and all associated components. The first area we call the `aperture area', and we define it as the area of the smallest convex polygon fully enclosing the space in which $|U(\vec{r})|>0$. The second area we call the `tile area', and we define it as the area of the smallest convex polygon fully enclosing all components associated with the radiating aperture (phase-shifters, splitter tree, etc.). The `aperture area' is the relevant metric for calculating aperture fill-factors, whereas the `tile area' is relevant for fabricating the tiled-array and the area needed to calculating TFF. For simplicity, here and in the main text we use the smallest rectangle enclosing the relevant space for calculating the aperture and tile areas.

To clarify the difference between the `aperture area' and `tile area', we consider using the single tile example case as the unit cell in the tiled-array example case. The `aperture area' we denote with the $tot$ subscript and for the single tile can be seen by inspection to be $A_{T,tot} = M\Lambda_TL$, the size of the rectangle enclosing all grating rows. The tiled aperture's `aperture area' is $A_{A,tot}=D_xN_xD_yN_y$ the rectangle enclosing all tiles. In this case the `tile area' of the single OPA tile is $D_xD_y$, the rectangle enclosing the tile, routing waveguides, and control elements. The tiled array also has a `tile area' which involves e.g. the splitter tree and control elements to distribute and control the individual tiles, which is not of interest here.

If we now apply the fill-factor equation to the example cases, we can calculate the single tile fill-factor as $\text{FF}_{T} = A_{T,eff}/A_{T,tot} = Mw_TL/M\Lambda_TL = w_T/\Lambda_T$ and the tiled-array fill-factor as $\text{FF}_{A} = A_{A,eff}/A_{A,tot} = N_xN_yA_{T,eff}/N_xN_yD_xD_y=A_{T,eff}/D_xD_y$. We use the fill-factor definition to identify $A_{T,eff} = A_{T,tot}(\text{FF}_T)$ and therefore $\text{FF}_{A} = A_{T,tot}/D_xD_y(\text{FF}_T)$. This substitution clarifies that the array fill-factor includes both the fill-factor of the tile $\text{FF}_T$ and the ratio of the tile's `aperture area' to its `tile area', $A_{T,tot}/D_xD_y$. The TFF defined in the text is exactly this ratio and is therefore agnostic to the tile design (fill-factor of the tile, $\text{FF}_{T}$). It can then be seen that the total fill-factor of the tiled-array is the TFF multiplied by the fill-factor of the tile $\text{FF}_{A} = (\text{TFF})(\text{FF}_{T})$, as expected.

We then identify that, whereas TFF is a key metric in determining the effects of tiling lobes, the effects of grating lobes are related to the single tile fill-factor FF$_T$ and the effects of both tiling and grating lobes are related to the total aperture fill-factor FF$_A$. It is therefore this total aperture fill-factor which is a good overall metric for determining the degradation in system performance due to unwanted lobes (grating and tiling) by fully accounting for both.

\subsection{Tiling Lobe Suppression}
\label{subsec:tiling_lobe_suppress}

The detrimental effects of tiling lobes, signal ambiguity and power loss, can be partially compensated with appropriate design. While the power lost to these lobes cannot be recovered without improving the fill-factor, it is possible to address the issue of signal ambiguity without changing the fill-factor. The approaches developed for grating lobe suppression (see Sec.~\ref{sec:grating_lobe_suppress}) such as aperiodic arrays \cite{hutchison2016high,komljenovic2017sparse,sayyah2015two} and Vernier lobe spacing \cite{pinna2018vernier,dostart2019vernier}, are equally applicable to tiling lobe suppression. For example, varying the tile periodicity between transmit and receive tiled-apertures can suppress tiling lobes, or even allow for sparse apertures which could be interleaved into a single, composite transceiver aperture.

\subsection{SNR of a Tiled-Aperture}
\label{subsec:snr}

The effects of tiling on the SNR of a LIDAR system are discussed briefly here. In particular, we are interested in how the TFF affects the SNR. Notably the following derivation is exact for apertures composed of separable rect functions, but only a useful approximation (which still accurately captures the scaling relations) for realistic aperture functions. The detector current due to the return signal is proportional to the received power from the main lobe $P_{rec}$ for direct detection, $i^D_S\propto P_{rec}$, and the geometric mean of the received power and the local oscillator (LO) power $P_{LO}$ for heterodyne detection, $i^H_S\propto\sqrt{P_{rec}P_{LO}}$ \cite{osche2002optical}.

The SNR is the temporally averaged ratio of electrical signal power $i^2_S$ to electrical noise power $i^2_N$

\begin{equation}
    \text{SNR} = 10\log_{10}\left(\frac{\langle i^2_S\rangle}{\langle i^2_N\rangle}\right)
\end{equation}

\noindent where the temporal averaging is denoted by $\langle\rangle$. Therefore the SNR is proportional to the square of the received power for direct detection $\text{SNR}^D\propto 20\log_{10}(P_{rec})$ or linear in the received power for heterodyne detection $\text{SNR}^H\propto 10\log_{10}(P_{rec}P_{LO})$ \cite{osche2002optical}.

In order to calculate $P_{rec}$, we require a `capture efficiency' $\eta_C$ which relates the radiated power $P_{rad}$ to the received power such that $P_{rec} = \eta_CP_{rad}$. This capture efficiency can be found by overlapping the field scattered off the target and receiver's back-projected field at a plane of our choosing, such as the target. If both fields are normalized ($\iint |U|^2dA=1$), we can write

\begin{equation}
    \eta_C = \left(\iint_\text{Main Lobe} U_{TX}(x,y)U_{RX}(x,y)R(x,y)dxdy\right)^2
\end{equation}

\noindent for a field reflectivity profile $R$, transmitter field on the target plane $U_{TX}$, and receiver back-projected field on the target plane $U_{RX}$. Note that the receiver field is \emph{not} conjugated as would be expected in an overlap integral (inner product) because the back-propagation operation (which is a time-reversal operation) applies a second conjugation cancelling out the conjugation from the overlap integral. With the notation here, $U_{RX}$ should be understood as the field on the target plane which would be generated if the receiver array was used as a transmitter.

We consider the case of both transmitter and receiver using identical OPAs. For this case it is easy to distinguish that the received power due to the signal (main lobe) is simply the overlap of the (identical) main lobes of both the transmitter and receiver and the integration term is $U_{TX}U_{RX}R\rightarrow U^2_{TX}R$. For an example array of $N_x\times N_y$ rectangular tiles with aperture size $w_x\times w_y$ at periodicity $D_x\times D_y$ the TFF is $w_xw_y/D_xD_y$ and the normalized field pattern on a target in the far-field at distance $z$ is

\begin{equation}
\label{eq:field_pattern}
\begin{split}
    U_{TX} = &\sqrt{\text{TFF}\frac{N_xN_yD_xD_y}{\lambda^2 z^2}}\\
    &\text{sinc}\left(\frac{N_xD_xx}{\lambda z}\right)  \ast_x\left[\text{sinc}\left(\frac{w_xx}{\lambda z}\right)\text{comb}\left(\frac{D_xx}{\lambda z}\right)\right] \\
    &\text{sinc}\left(\frac{N_yD_yy}{\lambda z}\right)\ast_y\left[\text{sinc}\left(\frac{w_yy}{\lambda z}\right)\text{comb}\left(\frac{D_yy}{\lambda z}\right)\right]
\end{split}
\end{equation}

\noindent where we have pulled the quadratic phase factor into the reflection phase and $\ast_x$, $\ast_y$ denote 1D spatial convolutions. For a diffuse reflector the capture efficiency is proportional to the field overlap $\eta_C\propto(\iint U^2_{TX}dA)^2$ (as well as aperture size and range) which, integrating over only the main lobe centered at $(x,y)=(0,0)$, can be found as $\eta_C\propto\text{TFF}^2$. For the case where only the transmitter is an OPA, it can be seen that $\eta_C$ includes only a single factor of TFF.

For identical transmit and receive OPAs we can therefore see that the detected signal power is proportional to the fourth power of TFF for direct detection and the square of TFF for heterodyne detection:

\begin{equation}
    \text{SNR}^D\propto \text{TFF}^4P^2_{rad}
\end{equation}
\begin{equation}
    \text{SNR}^H\propto\text{TFF}^2P_{rad}P_{LO}.
\end{equation}

\noindent The `tiling penalty' discussed in the main text is then the factor by which the SNR is degraded due to power loss, such that large tiling penalties correspond to low SNRs. For identical transmit and receive OPAs the tiling penalty TP is

\begin{equation}
    \text{TP}^D\propto \frac{1}{\text{TFF}^4}
\end{equation}
\begin{equation}
    \text{TP}^H\propto \frac{1}{\text{TFF}^2}.
\end{equation}

For the case where only the transmitter uses an OPA and the receiver is simply a large area, wideband detector behind a focusing telescope, these correspond to factors of 1/TFF$^2$ and 1/TFF respectively.

\subsection{Ambiguity}
\label{subsec:ambiguity}

In our SNR calculation we did not count the effects of tiling lobes as noise because erroneous signal due to tiling lobes is, in a sense, worse than noise. There are many signal processing methods for increasing SNR by boosting the signal power relative to the noise power. As one example, coherently integrating over multiple return pulses, which will be correlated in their arrival time, will suppress the noise relative to the signal proportional to the number of pulses. However, the tiling lobes are indistinguishable in all aspects from the main lobe post-detection and therefore these methods cannot be applied to improve the ratio of signal to erroneous signal. Because this erroneous signal is not noise but must still be accounted for, we introduce the ambiguity metric (Amb) which is the ratio of the erroneous signal power to the signal power. We define this metric along the lines of SNR: in terms of electrical signal power. The ambiguity is then

\begin{equation}
    \text{Amb} = \frac{\sum_i\langle i^2_{L,i}\rangle}{\langle i^2_S\rangle}
\end{equation}

\noindent the sum over the signal powers of all tiling lobes $i^2_{L,i}$ divided by the desired signal power. For this definition, $\text{Amb}\gg1$ corresponds to high ambiguity where the desired signal is indistinguishable from the erroneous signal, whereas $\text{Amb}\ll1$ denotes no ambiguity and the desired signal is significantly stronger than the erroneous signal.

The return signal due to a tiling lobe can be found identically with the main lobe case, and we therefore only require to calculate the sum over the squared of tiling lobe powers (for direct detection) or sum over the tiling lobe powers (for heterodyne detection). The sum over powers, in both cases, is not analytically tractable for our example array but can be approximated for both low and high TFF cases as

\begin{equation}
    \sum_{i\neq 0} P^2_{L,i} \propto 
    \begin{cases}
    \text{TFF}^4\left(1-\text{TFF}\right)^{10} &\text{TFF} >0.5\\
    \text{TFF}^4\left(\frac{0.2298}{\text{TFF}}-1\right) &\text{TFF} <0.15
    \end{cases}
\end{equation}

\noindent for direct detection and

\begin{equation}
    \sum_{i\neq 0} P_{L,i} \propto
    \begin{cases}
    \text{TFF}^2\left(1-\text{TFF}\right)^4 &\text{TFF} >0.6\\
    \text{TFF}^2\left(\frac{4}{9\text{TFF}}-1\right) &\text{TFF} <0.3
    \end{cases}
\end{equation}

\noindent for heterodyne detection.

\begin{figure}[t!]
	\centering
	\includegraphics[width=.8\textwidth]{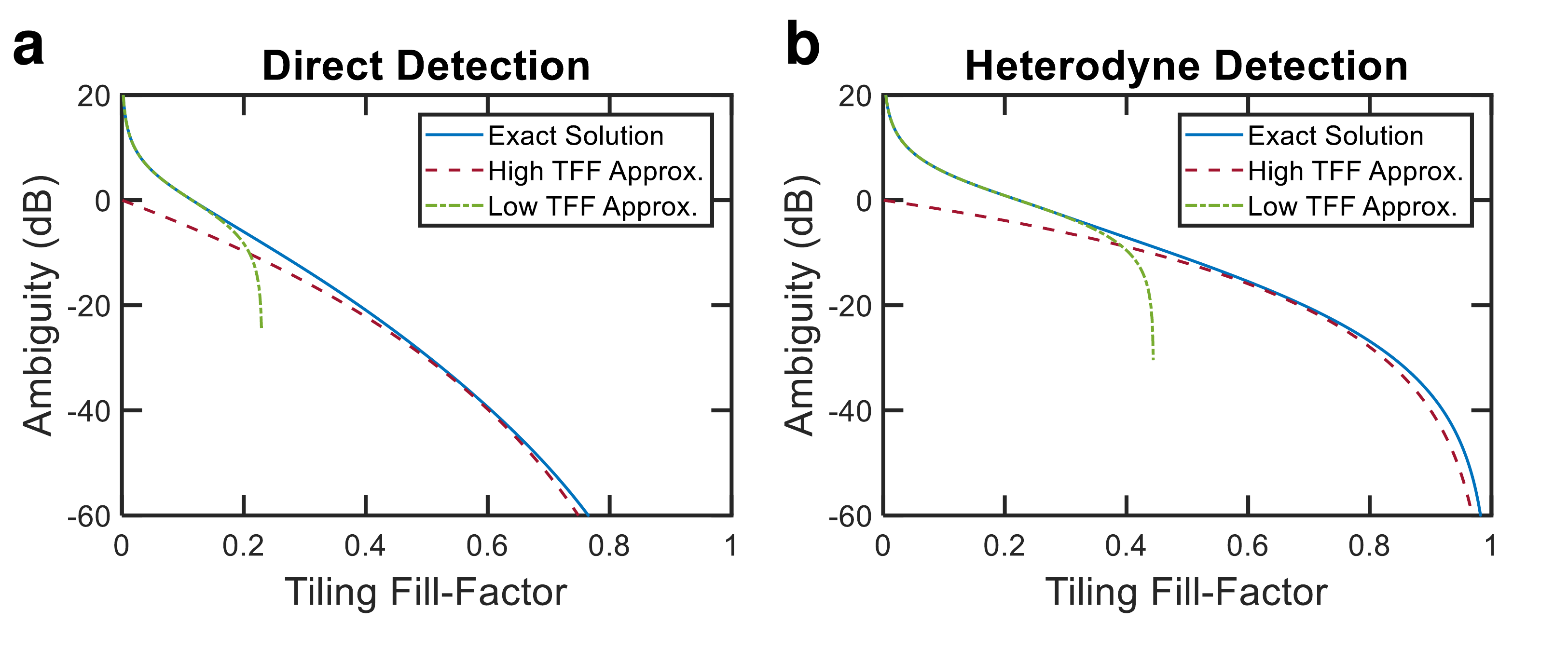}
	\caption{\textbf{Signal ambiguity due to tiling lobes.} \textbf{a} Signal ambiguity due to tiling lobes for a direct detection system. The exact (numerical) solution is plotted against the high and low TFF approximations. \textbf{b} Signal ambiguity due to tiling lobes for a heterodyne detection system with exact, high TFF, and low TFF solutions.}
	\label{fig:ambiguity}
\end{figure}

The ambiguities in the direct detection and heterodyne cases (where the other scaling factors cancel and therefore we avoid writing proportionality) are then

\begin{equation}
    \text{Amb}^D = 
    \begin{cases}
    \left(1-\text{TFF}\right)^{10} &\text{TFF} >0.5\\
    \frac{0.2298}{\text{TFF}}-1 &\text{TFF} <0.15
    \end{cases}
\end{equation}

\begin{equation}
    \text{Amb}^H = 
    \begin{cases}
    \left(1-\text{TFF}\right)^4 &\text{TFF} >0.6\\
    \frac{4}{9\text{TFF}}-1 &\text{TFF} <0.3
    \end{cases}
\end{equation}

\noindent where in both cases ambiguity grows inversely with TFF for low TFF, but approaches 0 as TFF approaches unity. The exact solution, as well as high TFF and low TFF approximations, are plotted for both direct detection and heterodyne detection systems in Fig.~\ref{fig:ambiguity}. Notably, while SNR was reduced by TFF more strongly for direct detection than heterodyne detection, the ambiguity of direct detection is more immune to TFF than heterodyne detection for TFF near unity by virtue of the stronger signal dependence on received power.

An ambiguity greater than 1 makes the system effectively useless except for specific LIDAR applications where there is no potential for reflection of grating lobes to create ambiguity; for example, observing spacecraft against a background of empty space. While SNR$\ll1$ is common, post-detection signal processing recovers this low SNR, resulting in an SNR$>1$ after signal processing that allows these systems to form useful LIDAR returns. Because there is not `signal processing' which allows for distinguishing the main lobe returns from tiling (or grating) lobe returns, we require an ambiguity less than 1 to allows the signal to be distinguished from other lobes. For direct and heterodyne detection, this limit corresponds to TFFs of 12\% and 22\%, respectively. Tiled apertures with TFFs below these limits will, on average, not be able to distinguish the main lobe signal from tiling lobe signals unless an ambiguity suppression approach is used, as discussed below. Only very few OPA designs, such as the SOPA, currently meet this requirement as seen in Table 1 in the main text. The presence of grating lobes will only worsen the issue, increasing the TFF required for unambiguous returns.

\section{Grating Lobe Suppression}
\label{sec:grating_lobe_suppress}

There are two main approaches to suppressing the effects of grating lobes: removing the presences of such lobes entirely by aperiodic spacing of emitters or misaligning the received grating lobes from the transmitted lobes. The first approach is generally referred to as aperiodic or random arrays, while the second (which uses periodic spacings) is referred to as Verniered arrays.

Aperiodic (or random) arrays \cite{hutchison2016high,komljenovic2017sparse,sayyah2015two} can be applied independently to transmitters and receivers and this approach spreads the unwanted lobe power out uniformly for a constant `background' while maintaining a diffraction-limited main lobe. The same amount of power is lost to the additional lobes, but the erroneous signal due to these lobes is suppressed and the ambiguity is removed. This approach necessitates spacing the gratings in an aperiodic manner, and will therefore slightly lower the fill-factor.

The Vernier approach \cite{pinna2018vernier,dostart2019vernier} requires designing the transmitter and receiver OPAs together such that they have different grating spacing periods. The different periods cause the additional lobes radiated by the transmitter to be misaligned with the equivalent lobes captured by the receiver, thereby suppressing the return signal. Just as with aperiodic approaches, the ambiguity is reduced at the cost of a slight decrease in fill-factor.

\section{Speckle Suppression Through Tile Current Summing}
\label{sec:current_summing}

When using OPAs as receivers in a LIDAR system it should be noted that OPAs suffer from speckle more than a standard incoherent detector due to the coherent capture of the back-scattered field, analogous to the heterodyne/mixing efficiency of a heterodyne bulk optical LIDAR even if used for a direct detection LIDAR. Specifically, for an aperture which contains $M$ speckles of the back-scattered field across the single OPA aperture, a heterodyne system or OPA will only convert one speckle's power to electrical signal on average due to the integral over the random phases of the speckle field. In contrast, an incoherent detector of the same size integrates over intensity and therefore converts a factor of $M$ more incident optical power to signal power (with variance of this speckled signal sensitivity increasing as $\sqrt{M}$) \cite{osche2002optical}.

For a transmitter of a given size projecting a beam onto a target at least as large as the beam spot, the correlation function of the speckle is given by the autocorrelation of the transmit aperture. Therefore, the speckle size at the receiver (in the same plane as the transmitter) is approximately twice the size of the transmitter. Therefore we optimally should build a receiver OPA half as large as the transmitter OPA, but no larger than the transmitter OPA, because there will be no benefit in terms of signal power from larger receivers. This motivation for a matched receiver can be inferred from the far-field overlap integral of the transmitted field and back-propagated capture field, which is maximized for identical transmit and capture fields.

It is possible to exceed this limitation by using a tile current summing approach, where each tile is coupled to its own detector and the detector currents are summed using a delay-matched tree to form the accumulated signal from the array of tiles. This contrasts with the more obvious approach of making the receive array identical to a transmit array: the laser feeding an array of tiles via a distribution network becomes an array of tiles which coherently combine their captured optical fields via a distribution network which feeds into a single detector. The benefit of the tile current summing approach is that, because each tile is smaller than a speckle, we effectively perform an incoherent integration at the tiled-aperture level and avoid the loss associated with coherent integration. The main drawback of this approach is that we cannot obtain any resolution enhancement beyond a single tile spot on the receiver end: we rely entirely on the tiled-aperture transmitter to provide the increased resolution. This effectively decouples the relative sizes of receiver and transmitter in terms of speckled return by making the receiver an incoherent detector on the scale of the speckle. 

\section{Limitations on Ranging with Wavelength-Steered OPAs}
\label{sec:ranging}

LIDAR is the main application intended for the SOPA demonstrated here, so some additional ranging restrictions imposed by wavelength steering should be noted. Consider a LIDAR system consisting of a CW laser and a modulator with RF bandwidth $B$ which will be used to modulate and encode the CW laser emission for time-of-flight ranging. After correlation of the return, the shortest correlation peak has a pulse width of $\tau\approx1/B$, such that higher bandwidth modulators can create shorter pulses. The temporal pulse width can be directly equated with a free-space pulse length as $l=c\tau$, the spatial extent of the pulse along the propagation axis.

The ranging operation itself involves detecting a pulse or coded waveform `echo' reflected off the target, measuring the delay $\Delta t$ between the emitted and detected waveform, and estimating the target range as $R=c\Delta t/2$. In the `many photon' regime, a large number of photons are detected each with their own measured time delay (or equivalently a classical, high speed detector) and the pulse shape can be reconstructed accurately. In this case the center of the pulse can be located well within the pulse width $\tau$, corresponding to a range resolution well below the pulse length $l/2$. However, LIDAR generally operates in a photon-starved (few photon) regime, where only several photons are detected per return pulse. In this case, it is common that the first photon is detected, but not later photons from the same pulse, as single photon detectors can have reset times much longer than the pulse width. Even if multiple photons are detected, noise in timing of the return pulse leads to a measurement uncertainty on the order of the pulse width. Thus the range resolution $\Delta R$ is given by $c\tau/2$, or in terms of bandwidth $\Delta R\approx c/2B$.

The SOPA, and other OPA approaches which utilize wavelength steering, map the optical frequency axis to a beam emission angle in 1D or 2D. Ranging uses the available bandwidth to determine the target range. The question therefore arises as to how to reconcile these two uses of the frequency domain and perform ranging in a wavelength-steered OPA. The answer lies in the disparity between the frequency ranges needed for wavelength-steering and ranging. For the predominant 1D wavelength-steered OPAs, 40-100 GHz frequency steps are used to steer a beam by one spot width, and only a few GHz of modulation is used for ranging, so the ranging signal does not cause noticeable beam deflection and these operations are effectively independent. For a ranging bandwidth on the order of (greater than) the beam steering frequency step (for example using picosecond laser pulses) then the frequency content of the pulse will cause beam deviations on the order of (greater than) the spot width. It can then be seen that a wavelength-steered OPA is only capable of projecting a bandwidth onto a single target spot at most equal to the spot steering frequency step, limiting the ranging resolution of a wavelength-steered OPA for a single resolvable spot.

In the case of the SOPA, as discussed in the main text, a 1.5 GHz step is required to steer by one resolvable spot along the fast axis. By design, this frequency step corresponds to a range resolution $\Delta R=10$ cm ($B=c/2\Delta R=1.5$ GHz), approximately the resolution desired for automotive applications. As mentioned in the main text, the minimum resolvable steering frequency step (corresponding to maximum ranging bandwidth) is determined by the total time delay across the aperture. For a wavelength-steered OPA, we can therefore determine the highest achievable range resolution $\Delta R$ in terms of the longest delay length $L$ (for a guided mode with group index $n_g$) as $\Delta R=Ln_g/2$. In an interesting equivalence, it can therefore be seen that the on-chip delay line length is of the same order, and linearly related to, the range resolution.

It should be noted that in the previous discussion we made no mention of the effects of the wavelength sweep range on ranging. This is because they are independent; the ranging bandwidth is entirely determined by the geometry (as is the spot width). Increasing the wavelength sweep range increases the FOV of a wavelength-steered OPA, and number of spots, but does not affect the spot size itself or the ranging bandwidth.

\bibliographystyle{ieeetr}
\bibliography{biblio.bib}